\documentclass{article}
\usepackage{geometry}                
\pdfoutput=1
\usepackage{graphicx}
\usepackage{natbib}
\usepackage{amsfonts}
\usepackage{amsmath, amssymb}
\usepackage{color}
\usepackage{ulem}

\newsavebox{\astrutbox}
\sbox{\astrutbox}{\rule[-5pt]{0pt}{20pt}}

\title{The Lagrangian description of aperiodic flows: a case study of the Kuroshio Current}

\author
{Carolina Mendoza$^{1,2}$, Ana M Mancho$^1$ \footnote{Email address for correspondence: a.m.mancho@icmat.es}\\
$^1$Instituto de Ciencias Matem\'aticas, CSIC-UAM-UC3M-UAM,\\ C/ Nicol\'as Cabrera 15, Campus Cantoblanco UAM, 28049 Madrid, Spain.\\
$^2$ETSI Navales, U. Polit\'ecnica de Madrid, Av. Arco de la Victoria 4, 28040 Madrid, Spain. }

\begin{document}

\maketitle

\begin{abstract}
This article reviews several recently developed Lagrangian tools and shows how their combined use    succeeds in  obtaining  a detailed description of purely advective transport events in general aperiodic flows.  In particular, because of the climate impact of ocean transport processes, we illustrate a 2D application on altimeter data sets over the area of the Kuroshio Current, although the proposed techniques are general
and applicable to arbitrary time dependent aperiodic flows. 
The first challenge  for describing transport in aperiodical time dependent flows is  obtaining a representation   of  the
 phase portrait where the most relevant dynamical features may be identified. 
 This representation is accomplished  by using global Lagrangian descriptors that when applied for instance to the altimeter data sets retrieve over the ocean surface a
 phase portrait   where the geometry of interconnected dynamical systems is visible. 
 The phase portrait picture  is essential  because it evinces 
 which transport routes are acting on the whole  flow. Once  these routes are roughly recognised it is possible to  complete a detailed description   by the direct computation 
of  the finite time stable and unstable manifolds  of  special  hyperbolic trajectories that act as organising centres of the flow. 
\end{abstract}

\section{Introduction}

The study of transport phenomena in aperiodic flows
is an important topic that arises in numerous applications. Lagrangian particle paths of non-periodic time dependent dynamical systems 
are the main ingredient of mixing processes, which take place in  manifold applications such as food production, microfluidics or geophysical flows. Mixing  is a key contributor to  significant features of the current climate when it takes place in the atmosphere or the oceans. 
 In the southern 
stratosphere, for instance, mixing across  the Antarctic polar vortex controls the springtime ozone depletion  \citep{alvaro,bernard}.  On natural catastrophes the pollutants mixing  in the ocean is also understood in terms of 
Lagrangian particle paths  \citep{mezic}. A better understanding of the mathematical tools describing transport in these contexts is important for improved control and prediction.

 Dynamical systems theory is the natural mathematical framework for describing
particle trajectories and transport in fluids where diffusion is not important. 
A challenge in the application of these tools to  realistic geophysical flows is that such  flows  are typically defined
as finite-time data sets and  are not periodic. An approach to these flows from a  geometric perspective includes the study of invariant manifolds, which act as barriers to particle transport and inhibit mixing. In this context   manifolds are approximated  by computing ridges of fields such as Finite Size
Lyapunov Exponents (FSLE) \cite{aur} and Finite Time Lyapunov Exponents
(FTLE) \cite{nese, slm05}. The latter authors  show that under certain conditions    there is small flux across FTLE  ridges.  
Despite the accomplishment of these techniques there exist frequent cases in which FTLE provide artifacts (see  \cite{bw10}) because  these assumptions 
are not satisfied.   Works such as \cite{bw10,mm11} have
noted 
ambiguity for the interpretation of these ridges and ambiguity over flow duration for FTLE calculations, in particular in transient flows. 
Another perspective within the geometrical approach different from Lyapunov exponents is that provided by distinguished hyperbolic trajectories (DHT) \citep{kayo,ju,chaos} and their stable
and unstable manifolds.  In this approach stable and unstable manifolds  are directly  computed  as material surfaces (see \cite{nlpg,mani}) thus  the flux across them is rigorously zero.
Distinguished trajectories are a generalization of the concept
of  fixed point for dynamical systems with a general time dependence.
 In this article we propose the use of DHT and their stable and unstable manifolds combined with recently developed {\it Lagrangian descriptors} \citep{chaos,prl,arxiv},
  which differ in some respects from other traditional techniques. 
  Our purpose is not confined to gathering/summarizing  these techniques in one article, but to providing a bigger picture that shows how the information they supply is
 complementary,  and that their combination constitutes a powerful package able for detecting the essential transport routes acting on an arbitrary flow.
For illustrative purposes we choose   a 2D application on  oceanic data.
In particular we consider  altimeter data sets over the area of the Kuroshio Current. Our choice is motivated by the fact that these data sets  are realistic and
obey no  regular pattern, as  might be objected to flows produced by exact analytical formulae.  The analysed flow is irregular and there is no  {\it a priori}  idea or control
on the transport mechanisms that take place on it.  We show that our tools are able to unveil the hidden dynamical picture of this arbitrary flow by tracing
medium and long term particle  transport  routes.  The performance  of the machinery on this  data opens a gateway to its  applications
  on any kind of realistic flow.

The Lagrangian analysis of altimeter data sets has been previously addressed by means of different approaches and for different purposes.
For instance \cite{ovidio1} have used  Finite-Size Lyapunov Exponents (FSLE) on  the geostrophic velocity field
to compute unstable manifolds which  are found to modulate phytoplankton fronts in lobular forms. 
\cite{ovidio2} have performed FSLE diagnosis on  altimeter data in the Algerian Basin, showing that Lyapunov exponents are able to predict the
(sub-)mesoscale filamentary processes not captured by an eulerian analysis. By computing probability distribution functions (PDFs) of the FTLEs over  currents derived from satellite altimetry,
  \cite{waugh} have evaluated 
global stirring variations.
\cite{josefina} have compared the lagrangian analysis provided by Finite-Time Lyapunov Exponents (FTLE)
on velocity fields obtained from two different multisatellite altimetry measurements, concluding that  both measurements
 support mixing with similar characteristics. In this context,  our work aims to extract transport routes in these realistic flows 
where there is no   {\it a priori}  idea on the transport mechanisms that take place on such flows.  We have chosen  the Kuroshio Current region for analysis 
in data measured during year 2003. We characterize transport events across a prominent jet and  
several eddies. Studies across these kinds of structures have been formerly discussed, either in {\it ad hoc} kinematic models \citep{samelson,grifa,meyers,duan,vulp}, or more recently
 in realistic flows \citep{roger,miller,kuz,jpo,bra}. Our purpose now is to show  how the combined
use of several recently developed Lagrangian tools,  valid for general time-dependent flows,  easily achieve insightful
transport mechanisms in this context.
 
The first step in our procedure  seeks for geometrical structures 
on the phase portrait (in advection it coincides with the physical space) where
 a sketch of the most relevant dynamical features may be identified at a glance. 
 This is achieved by means of a Lagrangian descriptor. Lagrangian descriptors are
based on a  recently defined function  \citep{prl}  which  evaluated over the vector field,
 succeed in covering the ocean surface with time-dependent geometrical
structures  that separate particle trajectories with different dynamical fates. 
  The organising centres of the flow are detected at a glance over the resulting map, and  the foliations induced by the stable and unstable manifolds of the present hyperbolic trajectories also are visible. We extend those results here and examine other possibilities for the definition of function $M$.
 The phase portrait picture indicates  transport routes  active on the    extended flow, and transport mechanisms
 such as the turnstile mechanism, where fluid interchange is mediated by  lobes, are sketched at this stage.
 Lobes may present a very tangled structure, especially in realistic flows such as the one under study, which makes  it very difficult  to 
 compute them accurately   from the representation  provided by the Lagrangian descriptor.  For this reason in order to
 proceed with a fine description of these pathways,  we then  characterize the organizing hyperbolic
 orbits and their stable and unstable manifolds by other techniques discussed in the literature  \citep{mani,nlpg,physrep,nlpg2}. 
  These methods are aimed at a direct computation of finite-time invariant manifolds. In the 2D dimensional case under study,  manifolds  are thus represented by lines 
 forming intricate lobes. From these clearly represented structures  complex particles paths may be traced out. 
  
  
The structure of the article is as follows. Section 2 provides a description of the dynamical system under study which is defined
from  altimeter datasets. These have been chosen to illustrate the use of these recent Lagrangian techniques in realistic flows. 
Section 3 discusses the role of Lagrangian descriptors as a first approach to this data. 
  Section 4 proceeds with the next step  where we explain how special trajectories that act as organizing centres of the flow are 
  characterised. We also explain how the direct computation of finite time manifolds is attained from them and discuss about frame invariance.
  Section 5 explains how manifolds   trace  complex and accurate transport routes, and abstract ideas from dynamical systems
  theory are shown to be present in  realistic datasets.  Finally section 6 presents the conclusions.

\section{The dynamical system}

\label{sec:altimetry}

We are interested in the study of transport on purely advective systems where particle evolution
is given by
\begin{eqnarray}
\frac{d{\bf x}}{dt} &=& {\bf v}({\bf x},t), \label{adv}  \, \,{\bf x} \in \, \mathbb{R}^n, \, t \in \, \mathbb{R}
\end{eqnarray}
In geophysical applications typically this expression takes $n=1, 2, 3$.
 We assume that ${\bf v}({\bf x},t)$ is  $C^r$ ($r \geq 1$) in ${\bf x}$ and continuous in $t$. This will allow for unique solutions
to exist, and also permit linearization, although linearization will not be used in our construction.
In our study, the velocity field  ${\bf v}$ given in Eq.  (\ref{adv}) is defined from observational data. We have 
considered a realistic 2D flow obtained from  altimeter data sets, with irregular time dependence far from periodicity. The fluid motion  involves  temporal transitions in which
eularian structures may be annihilated, created or move  rapidly. 
The flow is provided in a finite space-time grid, and our study will extract information assuming that data is 
well defined with the supplied resolution.  This means that below the scale of the grid the fluid behaves smoothly and it is well approached by 
a standard interpolation technique. There is no other a priori condition or hypothesis  on  it. 
Our purpose is to illustrate 
how recent  Lagrangian techniques  may be combined to approach a complete transport description on highly aperiodic dynamical systems.

 The velocity data set used  in this  work  has been previously described in \citep{turiel,nlpg2}, where many details are given. It has been processed  at  CLS Int Corp (www.cls.fr) in the framework of the
SURCOUF project \citep{larnicol}.  The data span the whole Earth, in the period from {November 20, 2002 to July 31,  2003}. Samples  are taken daily in a grid with $1080 \times 915$ points  which respectively correspond to longitude and latitude. 
The longitude is sampled uniformly from $0^o$ to $359.667^o$, however 
the Mercator projection is used between  latitudes $-82^o$ to $81.9746^o$ so this means that 
along this coordinate data are 
not uniformly spaced. The precision is $1/ 3$ degrees at the Equator.
Daily maps of surface currents combine altimetric sea surface heights and windstress data in a two-step procedure:
on the one hand, multimission (ERS-ENVISAT, TOPEX-JASON) altimetric maps of sea level anomaly (SLA) are 
added to the RIO05 global Mean Dynamic Topography \citep{Rio2,Rio4} to obtain global maps 
of sea surface heights from which surface geostrophic velocities ($u_g$, $v_g$) are obtained by simple derivation. 			
\begin{eqnarray}
u_g=-\frac{g}{f}\frac{\partial h}{\partial y} \\
v_g=\frac{g}{f}\frac{\partial h}{\partial x}
\end{eqnarray} 
where $g$ is  the gravitational constant and $f$ is the Coriolis parameter defined as follows:
\begin{eqnarray}
f= 2\Omega \sin(\lambda). \nonumber
\end{eqnarray} 
Here $\Omega=7.2921 \times 10^{?5} rad/s$ is the rotation rate of the Earth and $\lambda$ is the latitude.
On the other hand, the Ekman component of the ocean surface current (${\bf u}_{ek}=u_{ek}$, $v_{ek}$) is estimated using a 2-parameter  model :   
\begin{eqnarray}
{\bf u}_{ek}= b {\rm e}^{i\theta}\boldsymbol{\tau}. \nonumber
\end{eqnarray} 
where $b$ and $\theta $ are estimated by latitudinal bands from a least square fit between ECMWF 6-hourly windstress analysis   and $\boldsymbol{\tau}$ is an estimate of the Ekman current obtained removing the altimetric  geostrophic current from the total current measured by drifting buoy velocities available from 1993 to 2005.
The method  is further described  in \citep{Rio3}).
Both the geostrophic and the Ekman component of the ocean surface current are added to 
obtain estimates of the total ocean surface current. Despite the addition of  the Ekman component the resultant velocity  is almost divergence free, 
thus  motions are mainly  two dimensional. 
\begin{figure*}
\hspace*{-1.6cm}
\includegraphics[width=15cm]{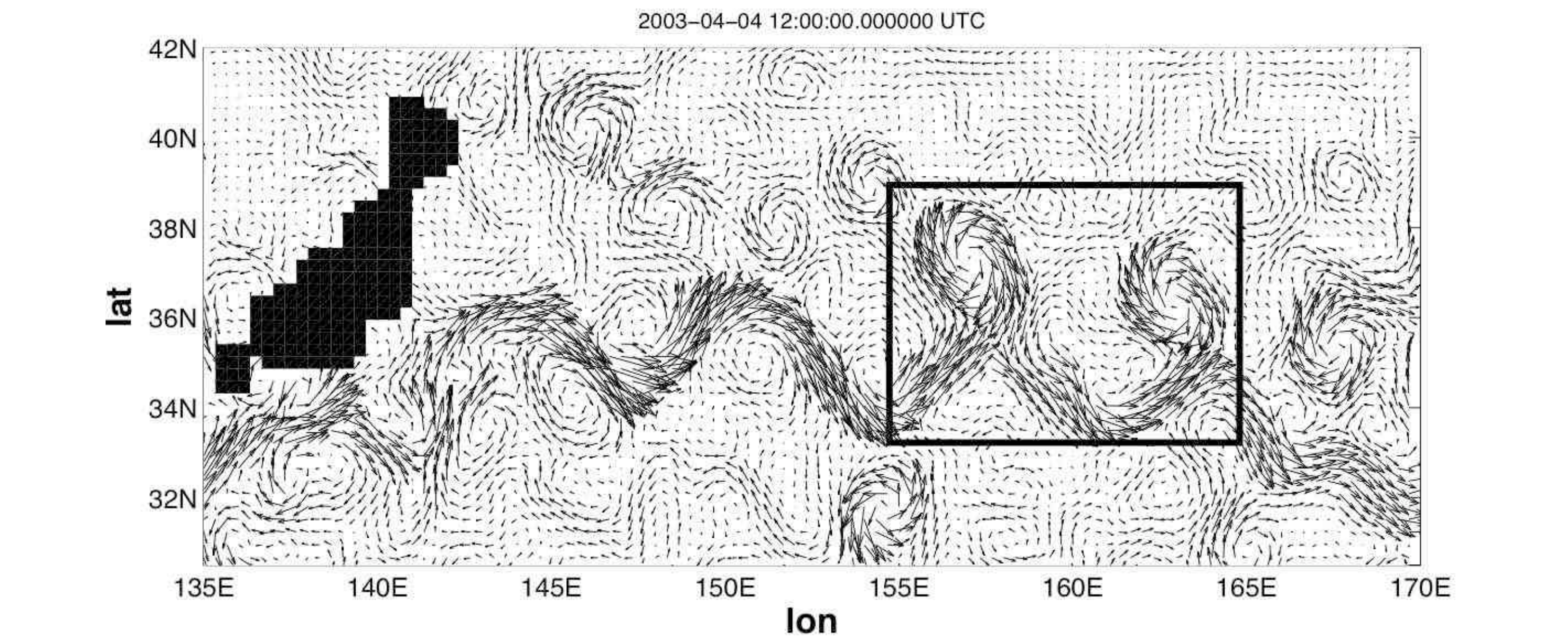}
\caption{\label{velocityfield} Velocity field of the Kuroshio current on April 4, 2003. The square highlights our main foucus area. Maximum
values of the velocity field are about { 3.65 m/s}. (Figure taken from \cite{nlpg2}). }
\end{figure*}

We focus over  a region through which the Kuroshio Current passes,  in 
April, May and June 2003.  A typical velocity field is shown in Figure \ref{velocityfield}. Our transport description is mainly focused on the region highlighted with a rectangle.
The equations of motion that describe the horizontal evolution of
particle trajectories on a sphere are
\begin{eqnarray}
\frac{d \phi}{d t}&=&\frac{u(\phi,\lambda,t)}{R {\rm cos}(\lambda)},\label{eqm1}\\
\frac{d \lambda}{d t}&=&\frac{v(\phi,\lambda,t)}{R}.\label{eqm2}
\end{eqnarray}
Here  the variables ($\phi, \lambda$) are longitude and latitude; $u$ and $v$ respectively represent the eastward and northward
components of the velocity field. 
The particle trajectories must be integrated
in equations (\ref{eqm1})-(\ref{eqm2}) and since information is provided solely  in a
discrete space-time grid, the first issue to deal with is that of
interpolation. We have daily maps of the velocity field and this is a coarse time grid  
to provide a time step in the integration of particle trajectories, however this frequency sampling is adequate in the sense that changes of the vector field below that resolution are smooth enough to be approached by an interpolator. 
Days are a typical time scale for the system (\ref{eqm1})-(\ref{eqm2}) and this is the unit of time in which results are reported.
A recent paper by \citet{msw}
compares different interpolation techniques in tracking particle
trajectories. Bicubic spatial interpolation in space \citep{nr}
and third order Lagrange polynomials in time are shown to provide
a computationally efficient and accurate method. We use this
technique in our calculations as it has been successfully implemented  in realistic flows over a sphere
as discussed in \citep{jpo}. Following this work   we notice that bicubic
spatial interpolation requires 
a uniformly spaced grid, while our data grid   is not uniformly spaced in the latitude
coordinate. We transform our coordinate system to a new
one ($ \phi, \mu$), in which the latitude
$\lambda$ is related to the new coordinate $\mu$ by
\begin{equation}
\mu=   {\rm ln}|{\rm sec}\lambda+ {\rm tan}\lambda| \label{mul}
\end{equation}
Our velocity field is now on a uniform grid in the ($\mu, \phi$)
coordinates. The equations of motion  in the new variables are,
\begin{eqnarray}
\frac{d \phi}{d t}&=&\frac{u(\phi,\mu,t)}{R \, {\rm cos}(\lambda(\mu))} \label{sd1}\\
\frac{d \mu}{d t}&=&\frac{v(\phi,\mu,t)}{R \,{\rm
cos}(\lambda(\mu))} \label{sd2}
\end{eqnarray}
In the numerical simulations   the vector field in Eqs.    (\ref{eqm1})-(\ref{eqm2}) is   represented in a selection of data 
spanning a domain in longitude and latitude $(\phi_{min},\phi_{max})\times(\lambda_{min},\lambda_{max})=(109.66^o,259.66^o)\times(14.74^o,59.56^o)$ which is much larger  than those displayed  in figures.
This ensures that particle integrations do not  cross the edges and thus boundary effects are not present. Variables $(\phi, \mu)$ are further transformed  by scaling the domain to $(0,3)\times(0,1)$,
which is more convenient for the manifolds computations reported in Section 4.2. The new variables are
\begin{eqnarray}
         x_1=3\frac{(\phi-\phi_{min})}{(\phi_{max}-\phi_{min})}\\
         x_2=\frac{(\mu-\mu_{min})}{(\mu_{max}-\mu_{min})}
\end{eqnarray}
The scaling provides the dynamical system in which integrations are performed:
\begin{eqnarray}
         \frac{d x_1}{dt}&=&v_1(x_1,x_2,t)\label{sd1def}\\
         \frac{d x_2}{dt}&=&v_2(x_1,x_2,t).\label{sd2def}
\end{eqnarray}
Once  trajectories are integrated  for
presentation purposes, one can convert coordinates back to the original ones. In the reversion $x_2\to\mu\to \lambda$ 
we use the expression $\lambda(\mu)$ obtained by inverting Eq. (\ref{mul}), i.e.
\begin{equation}
\lambda=\frac{\pi}{2}-2 \,{\rm arctan}(e^{-\mu}) \ . \label{lmu}
\end{equation}

\section{A time-depedent phase portrait}

\begin{figure*}
a)\includegraphics[width=9cm]{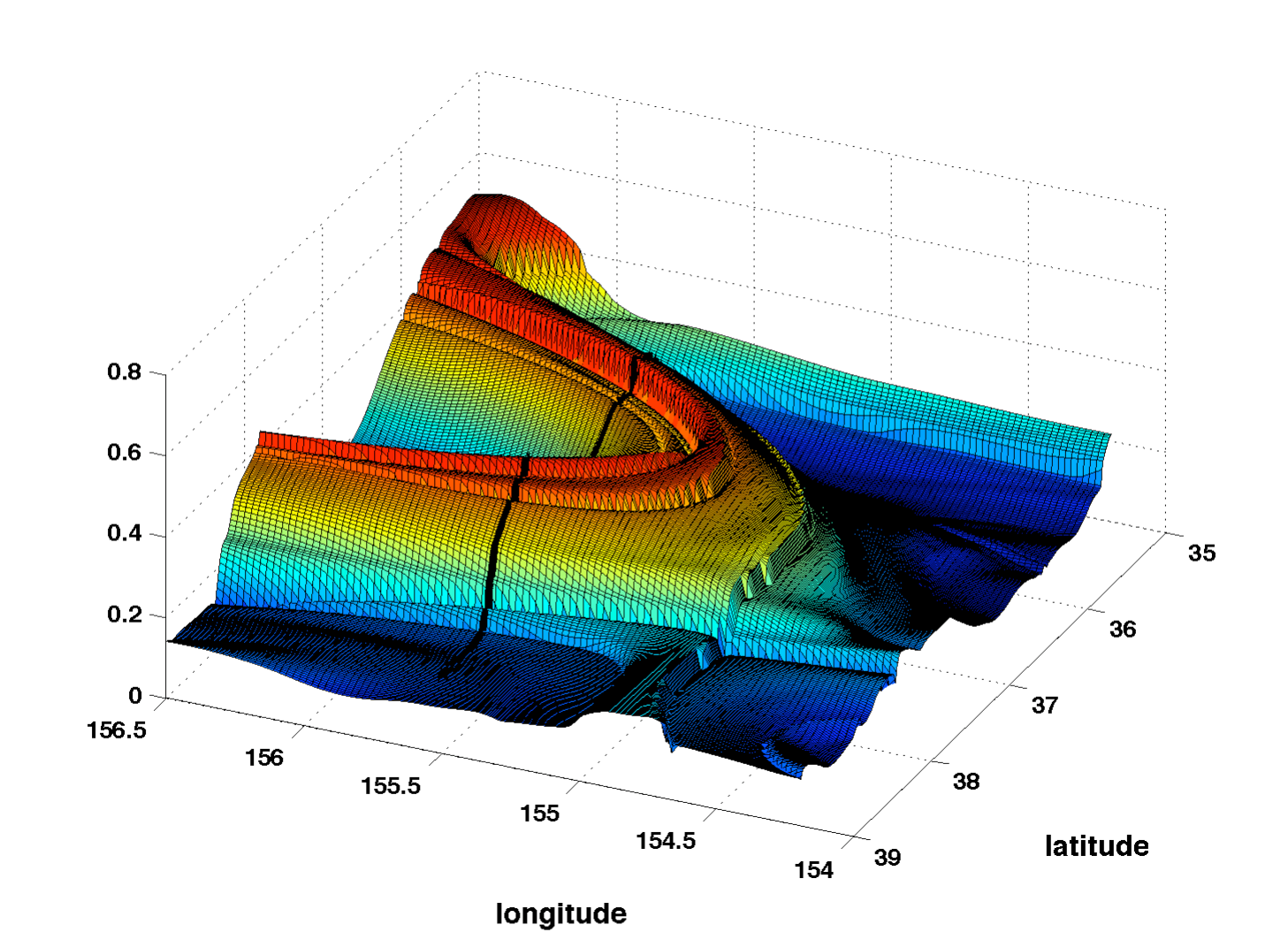}b)\includegraphics[width=6.5cm]{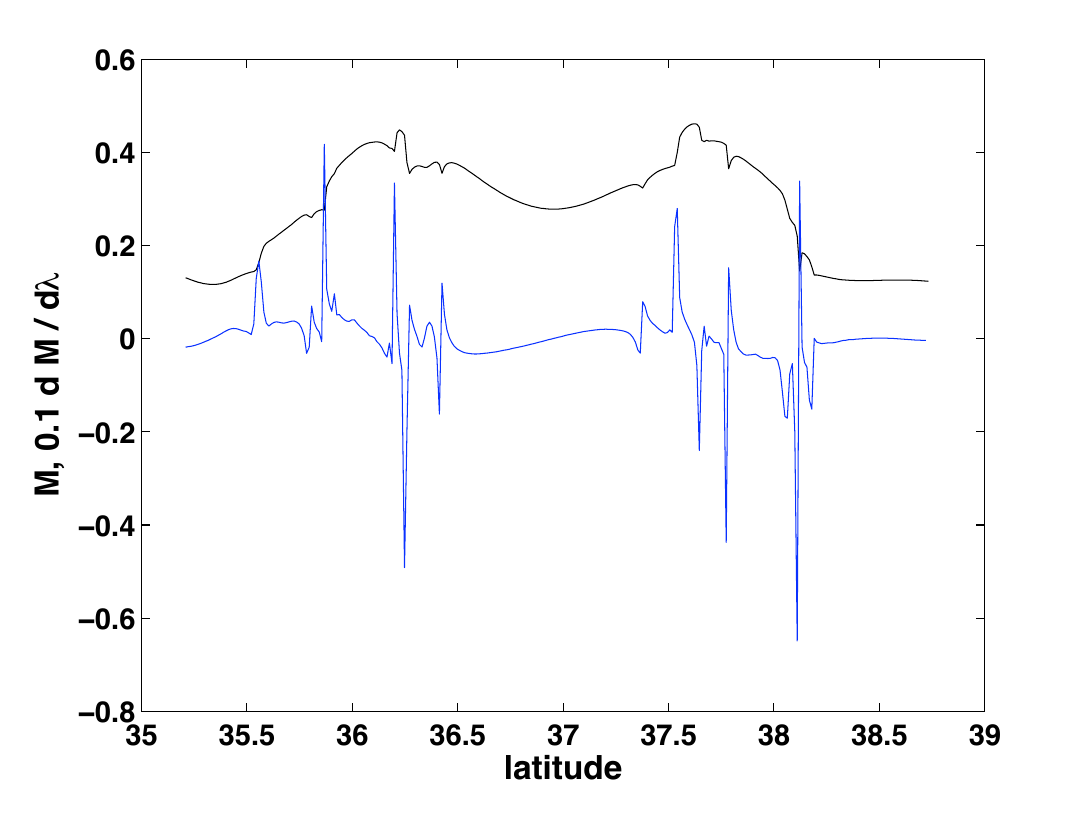}
\caption{\label{fig:M0} a) A representation of  the function $M$ over a small oceanic area on May 2, 2003 for $\tau=15$ days; b) In black the function $M$ vs latitude at a fixed longitude highlighted in a) with the thick black line. Abrupt changes in $M$ pointing manifolds positions corresponds with discontinuities on the derivative. In blue 0.1 times the derivative of $M$ with respect the latitude.   }
\end{figure*}
  \begin{figure*}
a)\includegraphics[width=5.5cm]{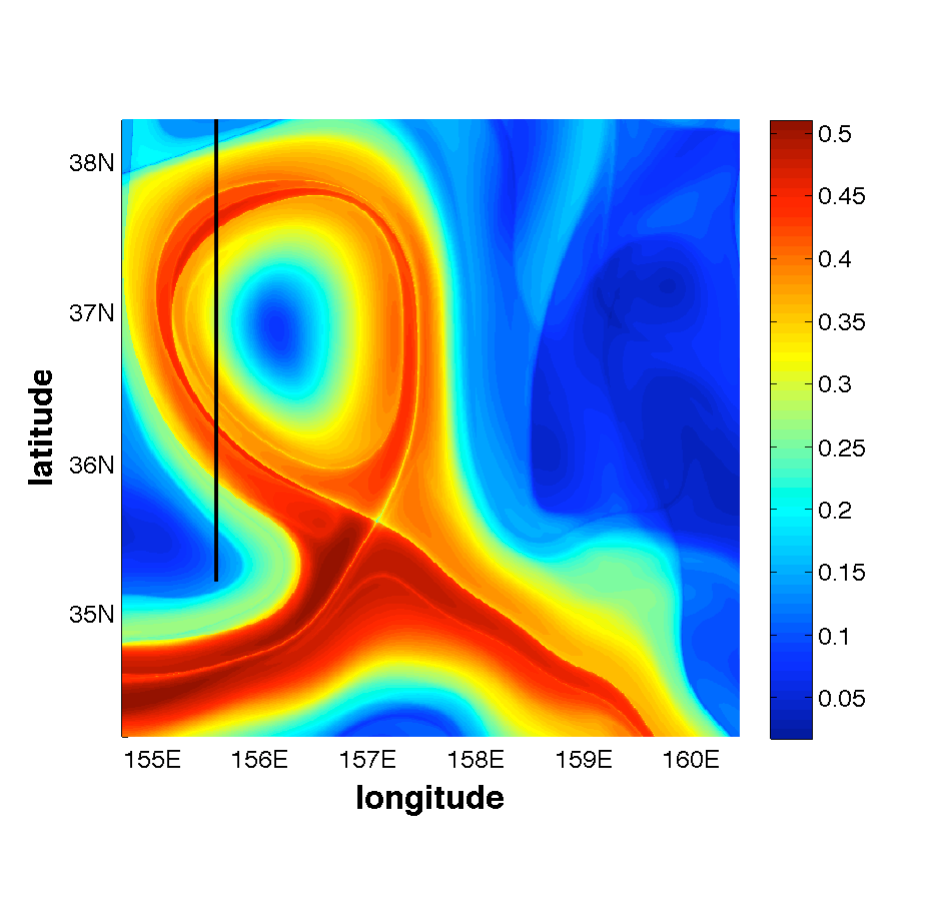}b)\includegraphics[width=7.5cm]{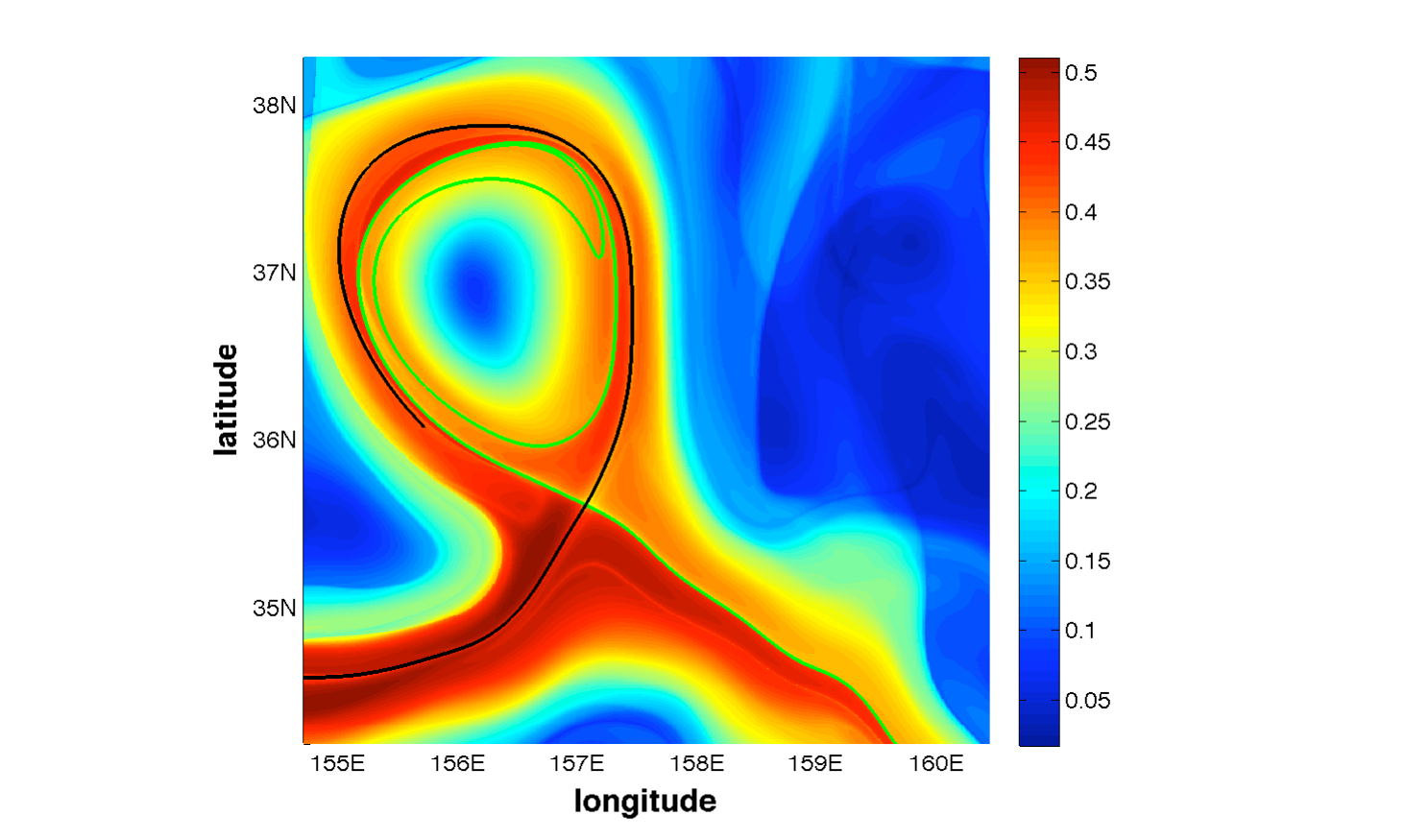}c)\includegraphics[width=5.5cm]{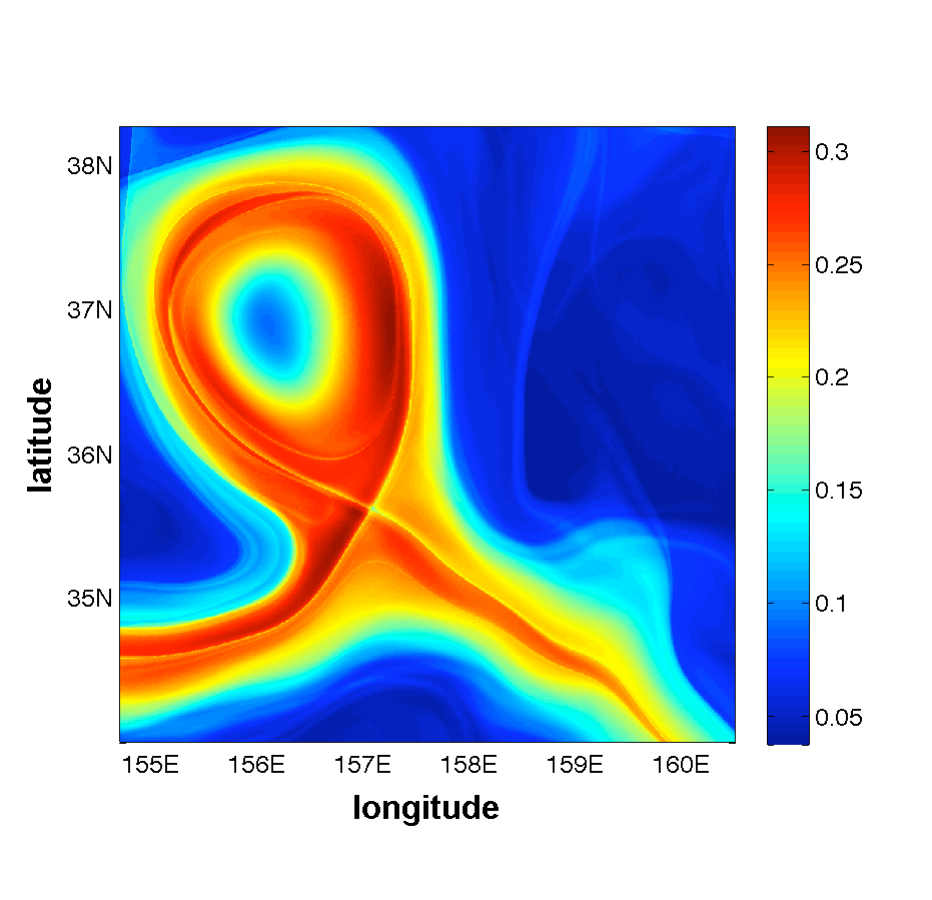}
\caption{\label{fig:M0b}  a) Contour plot of the   function $M$ over a small oceanic area on May 2, 2003 for $\tau=15$ days. The black straight line corresponds to the selection for outputs in Fig 2 a) ; b) the same with a piece of stable manifold (black line) and a piece of unstable manifold (green line) overlapping; c) the function $M$ for  $\mathcal{F}({\bf x}(t))=||{\bf a}({\bf x}(t),t)||$ and $\tau=15$ days in the same area.}
\end{figure*}

Solutions of dynamical systems are qualitatively described  according to Poincar\'e's idea of seeking geometrical structures 
on the phase portrait.  These can be used to organise particles schematically by regions corresponding 
to qualitatively different types of trajectories. In time independent systems --those in which Eq. (\ref{adv})
does not depend explicitly on time--   fixed points are   essential for describing  the solutions geometrically. 
Fixed points
may be classified as hyperbolic or non-hyperbolic depending on their stability
properties. Stable and unstable manifolds of hyperbolic fixed points 
act as separatrices that divide the phase portrait in regions in which particles have different dynamical fates. 
 To achieve this geometrical representation in time dependent aperiodic dynamical systems  is a challenge, because the concepts used in autonomous dynamical systems do no apply directly to these systems.
In these cases,   structures containing Lagrangian information on the time-evolution of fluid particles
 have typically been obtained by means of Lyapunov exponents.  The concept of Lyapunov exponent is infinite time  
and it is used in finite-time data sets  for its finite-time versions such as  finite-size Lyapunov exponents (FSLE) \citep{aur}
 and finite-time Lyapunov exponents (FTLE) \citep{haller,nese}. 
 
Different  Lagrangian tools that also succeed in  finding 
time dependent partitions  for finite time aperiodic geophysical flows are proposed in  this section. These implements are called  Lagrangian descriptors. 
Lagrangian descriptors provide a global dynamical picture of arbitrary time dependent flows by detecting simultaneously the organizing centers of the flow, 
hyperbolic trajectories and their stable and unstable manifolds and elliptic regions.  This technique has been successfully applied  by \cite{dcmism12} to stratospheric re-analysis data  produced by  the interim European Centre for Medium-Range Weather Forecasts (ECMWF) and has allowed the detection of dynamical features not perceived by other methods.  Originally Lagrangian descriptors 
 were  introduced by \cite{prl}, in the context of altimeter velocity data,  who proposed a function to this end.  This function  is referred to as $M$ and  was 
 advanced in \cite{chaos} as a building
block of the definition of {\it Distinguished trajectories}. These trajectories,  their organizing  role and  their computation from $M$ are discussed further in the next section. We now  focus on the 
capacity of $M$ as a Lagrangian descriptor. The function $M$  measures the 
Euclidean arc-length of the curve outlined by a trajectory passing through 
 ${\bf x^*}$ at time  ${ t^*}$ on the phase space. The trajectory is integrated from ${ t^*}-\tau$ to ${t^*}+\tau$. This is mathematically expressed as follows:
 For all initial conditions ${\bf x^*}$   in an open set ${\mathcal B}\in\mathbb{R}^n$, at a given time    ${ t^*}$,
the   Lagrangian descriptor is a function $M({\bf x^*}, t^*)_{{\bf v},\tau}:({\mathcal B}, t) \to\mathbb{R}$ given by: 
\begin{equation}
M({\bf x^*}, t^*)_{ {\bf v}, \tau}=   \int^{t^*+\tau}_{t^*-\tau} \! \!\!\sqrt{\sum_{i=1}^n \left(\frac{d x_i(t)}{dt}\right)^2 }dt  \label{def:Mgen}
\end{equation}
Here  $(x_1(t) ,x_2(t),...,x_n(t))$ are the components in $\mathbb{R}^n$ of a trajectory ${\bf x}(t)$.
  The function $M$ depends  on  $\tau$ and also on the vector field {\bf v}.  It is defined for dynamical systems in arbitrary dimension $n$, but  for the chosen system (\ref{sd1def})-(\ref{sd2def}),
  $n=2$. 
 The question is why should $M$
succeed in realizing Poincar\'e's idea,  revealing the geometry of objects such as the stable and unstable manifolds of hyperbolic trajectories?
\cite{prl} report this observed fact, and although it is not formally proven,  an heuristic argument on this evidence is given.
$M$ measures the arc-length of trajectories on a time interval $(t^*-\tau, t^*+\tau)$. 
For a given $\tau$ there may be trajectories that start and evolve close to each other and this of course may change with $\tau$. 
Trajectories which stay close are expected to have similar arc-lengths.  However  for this $\tau$, at the boundaries of  
regions  comprising trajectories with  qualitatively different evolutions, arc-lengths  will change abruptly, and these regions are exactly what the stable and unstable manifolds separate. 
We evaluate $M$ as defined in Eq. (\ref{def:Mgen}) over the oceanic velocity field and a first output is provided in  Figure \ref{fig:M0}. The coordinates at which sharp changes on $M$ occur are related to points of discontinuity on the derivative along a direction which is non tangent to the manifold. These  are disclosed in Figure \ref{fig:M0}b). 
A contour plot of  the same area,  portrayed in Figure \ref{fig:M0b}a),  links the positions for  these abrupt variations to lines resembling  singular features. Figure \ref{fig:M0b}b)
 visually demonstrates that the coordinates at which singular lines of the function $M$   are placed coincide with the positions of the stable and unstable manifolds.  The Lagrangian information provided by $M$, 
 that is the position of the invariant manifolds, is not contained on the specific values taken by  $M$ but on the positions at which  these values change abruptly. However in the interest of completeness, figures show a color bar indicating the range of $M$. Units correspond to those in the rescaled 
 system (\ref{sd1def})-(\ref{sd2def}).

  Eq.   (\ref{def:Mgen})  finds arc-lengths integrating the modulus of the velocity ($||{\bf v}||$) along a trajectory. It is easily observed that
 the heuristic argument  should in fact work for the accumulation of other
 positive  intrinsic geometrical or physical  property along  trajectories 
on a time interval $(t^*-\tau, t^*+\tau)$.  
For instance could have been considered integrations of  the modulus of acceleration ($||{\bf a}||$), the modulus of the time derivative of acceleration ($||d{\bf a}/dt||$), or positive scalars obtained from combinations
of ${\bf v}$, ${\bf a}$ or $d{\bf a}/dt$ as far as these combinations are bounded. 
In this way trajectories evolving close to each other during this time interval would accumulate a similar value for $M$,
and the  accumulated value of the property would  be expected to change sharply  at the boundaries of  
regions  comprising trajectories with  qualitatively different evolutions. These abrupt changes would  highlight  the stable and unstable manifolds.
Our general method for building up families of Lagrangian descriptors  for general time dependent flows replaces   Eq.   (\ref{def:Mgen}) by
\begin{equation}
M({\bf x^*},t^*)_{ {\bf v},\tau}=  \int^{t^*+\tau}_{t^*-\tau} \mathcal{F}({\bf x}(t)) \, dt. \label{def:M}
\end{equation}
where $\mathcal{F}({\bf x}(t))$ denotes a bounded positive intrinsic physical or geometrical property of the trajectory ${\bf x}(t)$. In practice not all choices of
$\mathcal{F}({\bf x}(t))$ are equivalent. Typically for analyzing velocities fields 
given as  data sets,
choices involving $||d{\bf a}/dt||$ may be less appropriate  than those
involving ${\bf v}$ or ${\bf a}$
because they require interpolators with a higher order of regularity  than the latter magnitudes. Similarly a choice involving  ${\bf a}$ requires  an interpolator with a higher regularity  than those involving only ${\bf v}$. In this section we report results for $\mathcal{F}=||{\bf v}||$ and $\mathcal{F}=||{\bf a}||$. Both choices are adequate for the type of interpolation used in the velocity field. Many other options on $\mathcal{F}$ are thoroughly discussed and compared 
in \citep{arxiv}. For comparison purposes Figure \ref{fig:M0b}c) shows the output obtained 
 when  $M$ is evaluated as in Eq. (\ref{def:M}) with the choice  $\mathcal{F}({\bf x}(t))=||{\bf a}({\bf x}(t),t)||$.  As anticipated,  singular lines in the contour plot  coincide with the position of invariant manifolds. Full details of the numerical evaluation of $M$ are given in the Appendix A.

The heuristic argument pointed out above, supports the ability of  Lagrangian descriptors for  highlighting manifolds,  but it is not a rigorous argument. The power of Lagrangian descriptors 
however is sustained by a strong numerical evidence consistently shown in all the examined examples, which thus inspires the development of further theoretical  results.  

\begin{figure}
\hspace{2cm}
\includegraphics[width=6.5cm]{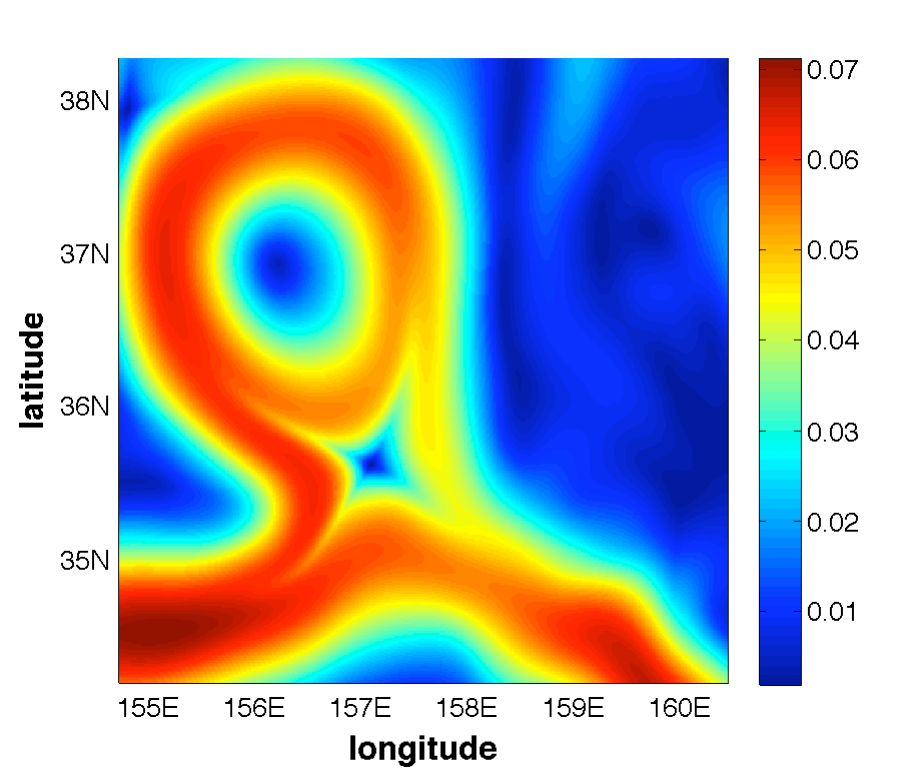}
\caption{\label{fig:M0c} A representation of  the function $M$ over a small oceanic area on May 2, 2003 for $\tau=2$ days. (Color version of a figure from \cite{prl})}
\end{figure}

\begin{figure*}
a)\includegraphics[width=8cm]{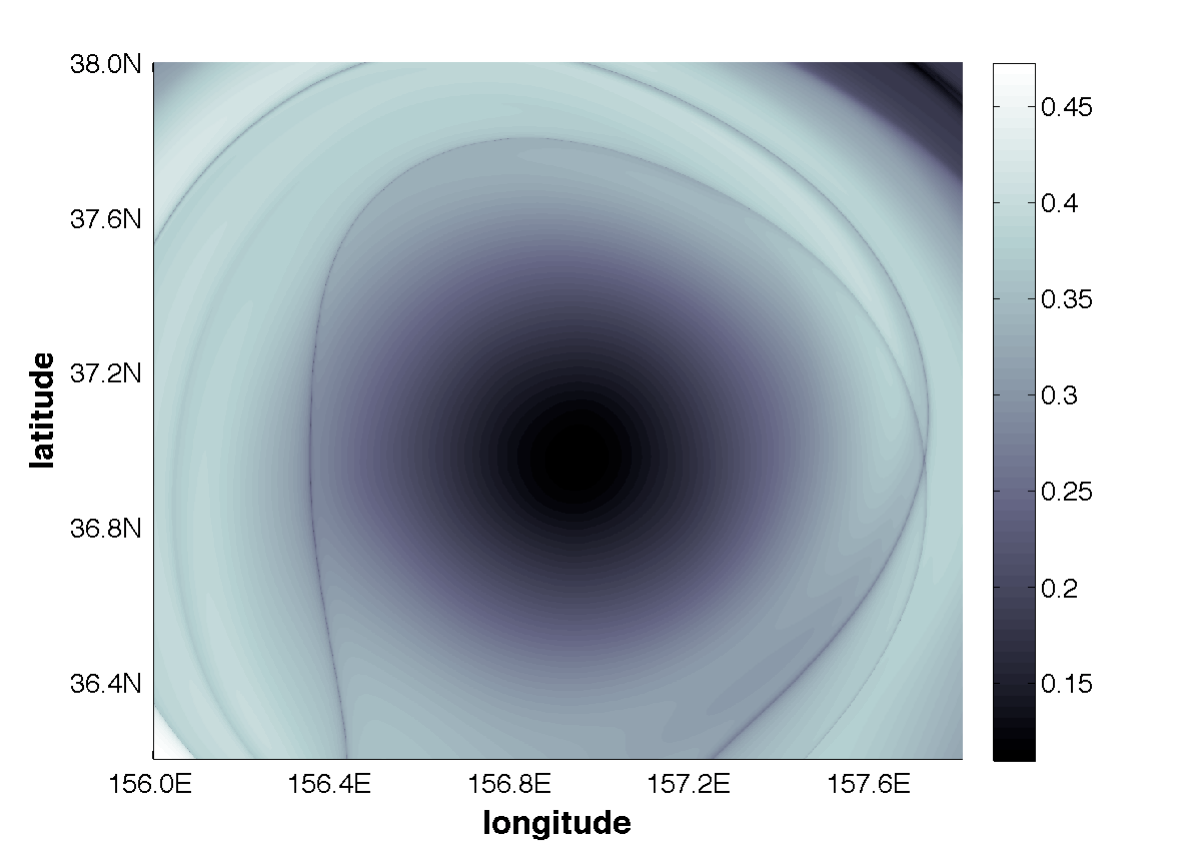}b)\includegraphics[width=8cm]{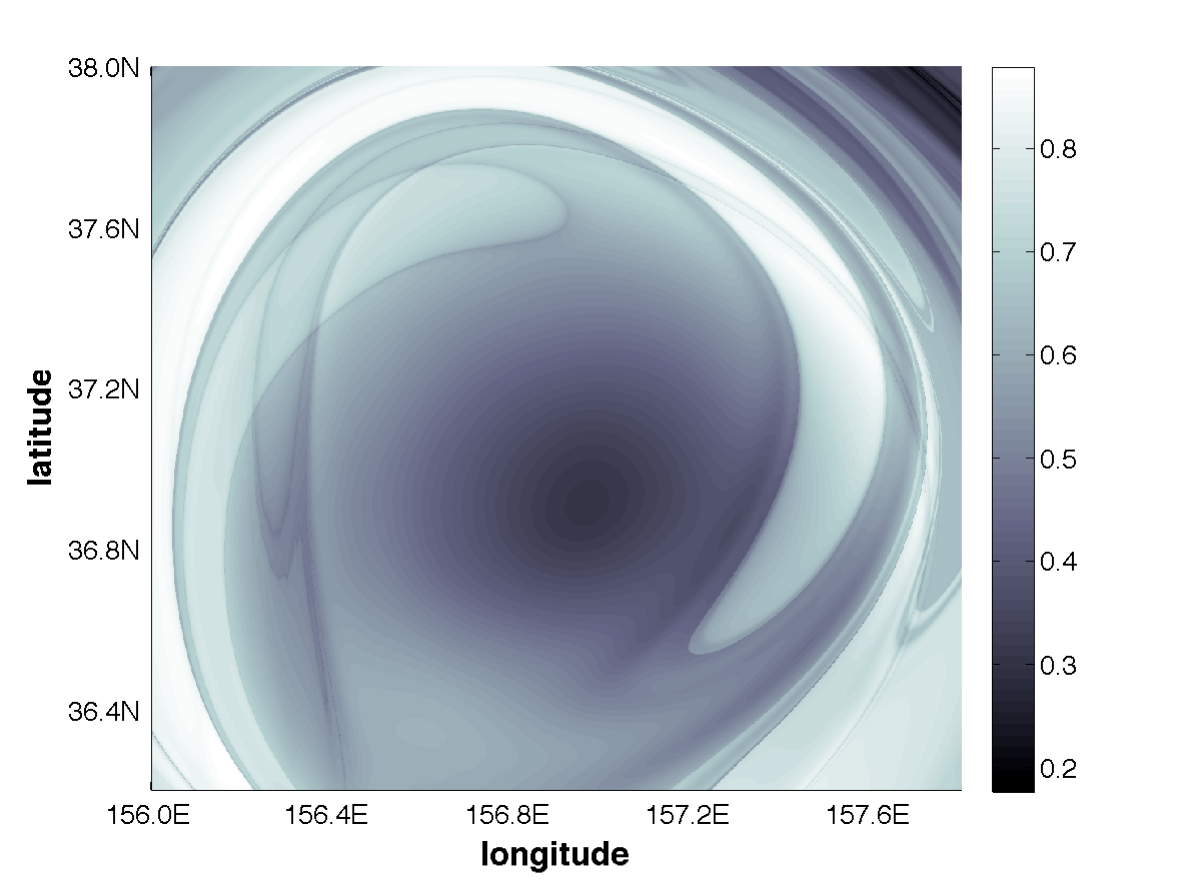}\\
c)\includegraphics[width=8.cm,height=6.3cm]{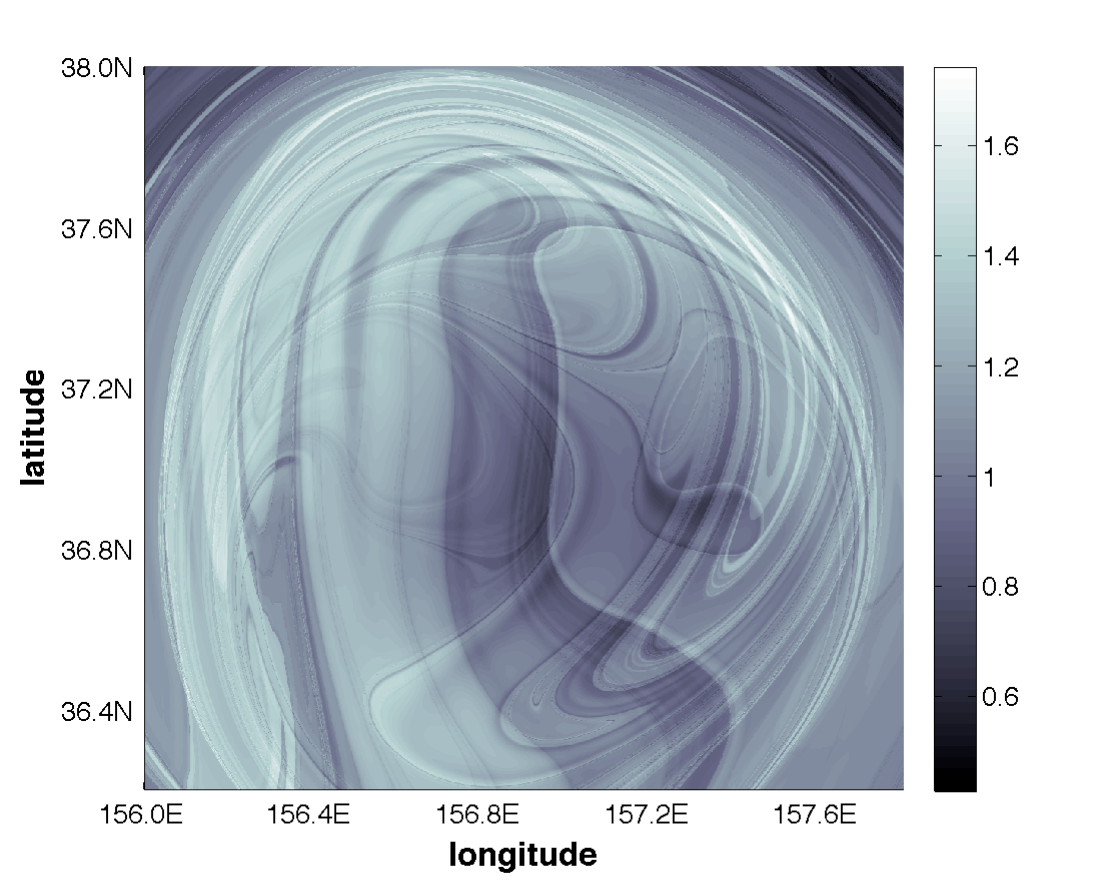}d)\includegraphics[width=8cm]{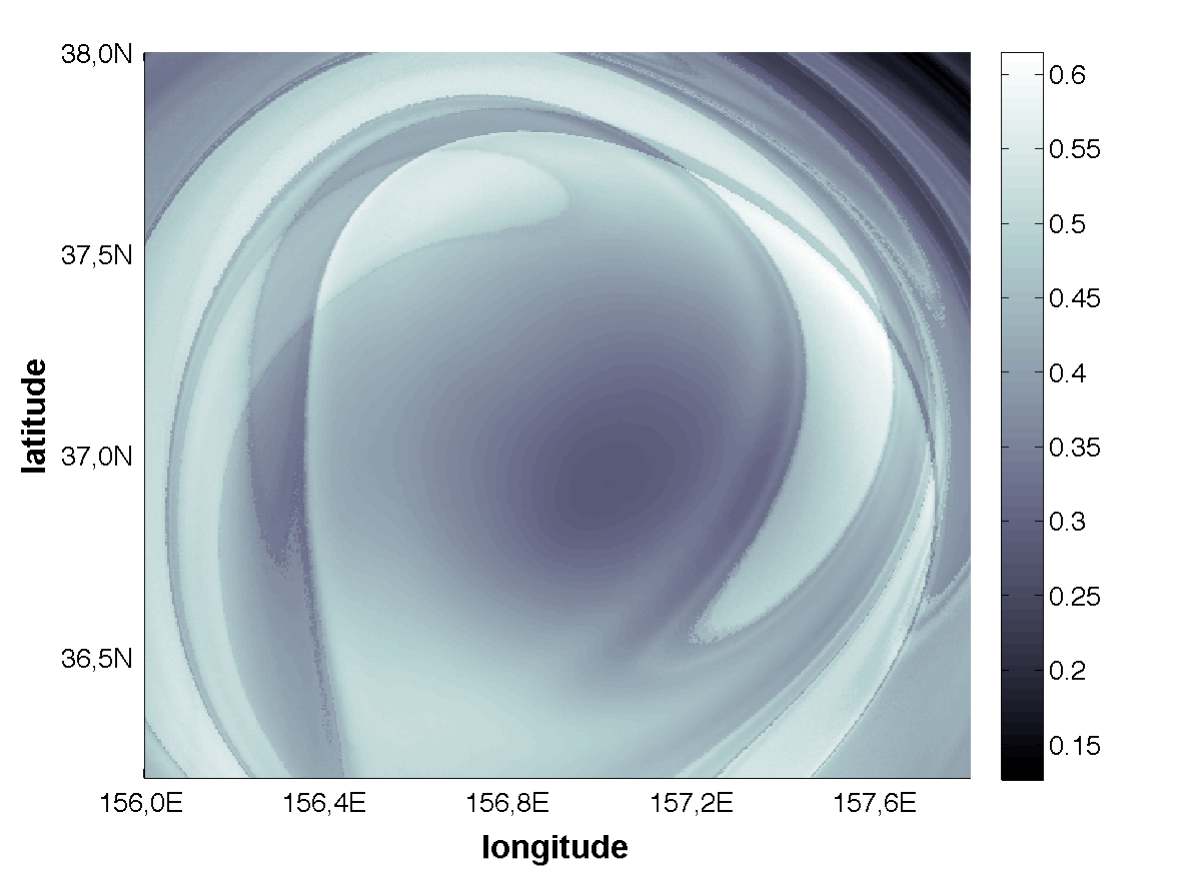}
\caption{\label{meddie} Lagrangian structure of the inner core of the western eddy on April 17 for increasing $\tau$ values. a)  $\mathcal{F}({\bf x}(t))=|{\bf v}({\bf x}(t),t)|$ and $\tau=15$ days; b)  $\mathcal{F}({\bf x}(t))=|{\bf v}({\bf x}(t),t)|$ and  $\tau=30$ days; c)  $\mathcal{F}({\bf x}(t))=|{\bf v}({\bf x}(t),t)|$ and  $\tau=72$ days; d)  $\mathcal{F}({\bf x}(t))=|{\bf a}({\bf x}(t),t)|$ and  $\tau=30$ days. }
\end{figure*}

\begin{figure}
\includegraphics[width=9.5cm]{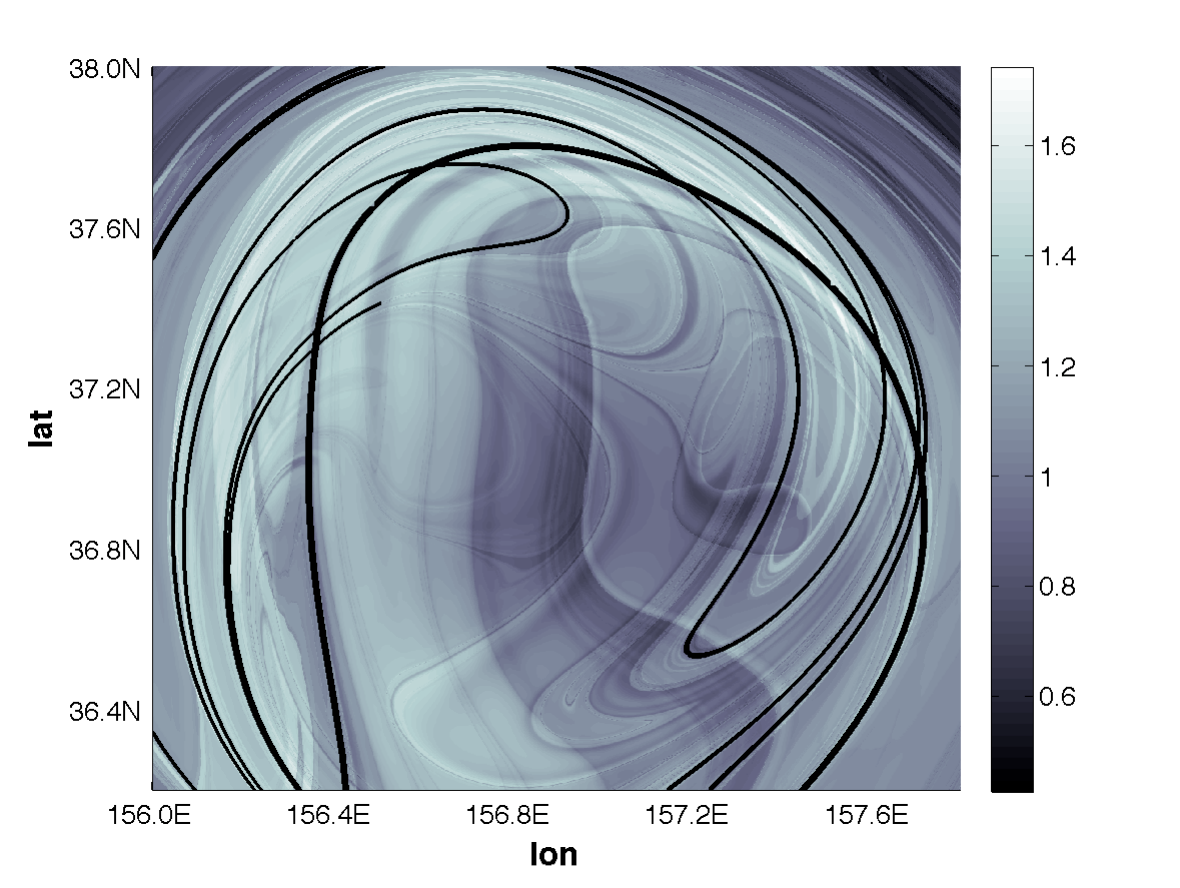}
\caption{\label{overlap} Stable and unstable  manifolds overlapped on the function $M$ at day April 17 for $\tau=72$ days. 
There is a coincidence between singular features of  $M$ and  manifolds.   }
\end{figure}

The function $M$ depends on $\tau$ in such a way that at low $\tau$, its structure is far from depicting manifolds.  For instance, for  $\tau=2$,
Figure \ref{fig:M0c} shows a contour plot of $M$ for  $\mathcal{F}({\bf x}(t))=||{\bf v}({\bf x}(t),t)||$,  at the same coordinates as in Figure \ref{fig:M0b}, but the observed structure is smooth and eulerian-like. The structure of $M$ at low $\tau$ is closely related to
the spatial  structure   of the velocity field, thus for highly turbulent flows with a more complex spatial structure,  $M$  is expected  to display a  richer pattern. 
 Figure  \ref{meddie} shows contour plots of $M$  on April 17 over an area with an eddy-like vector field. 
For increasing  $\tau$, $M$ displays more and more  complex patterns and outlines a growing manifold structure.
In Figures  \ref{meddie}a) and b), at low $\tau$ values the structure of $M$ at the inner part of the eddy has a minimum   which is locally smooth. 
This implies that  in the range $(t-\tau, t+\tau)$,
trajectories in this  vicinity outline similar paths: there are no sharp changes, and  thus they   behave as a coherent structure. 
The boundaries of this smooth region separate the mixing region (outside the core) from the non mixing region (inside).  A comparison between Figures   \ref{meddie}b) and   \ref{meddie}d) confirms that both descriptors report similar outputs.
In two-dimensional, incompressible, time-periodic velocity
fields, this kind of structure is typical because
 the KAM tori enclose the core --a region of bounded
fluid particle motions that do not mix with the surrounding
region \citep{wiggins}. However, there is no KAM theorem for velocity fields
 with a general time-dependence \citep{wiggsamelson} such as the one in our analysis.
 In this context, a  question  that remains open is to address the dispersion or confinement 
of particles in the core for aperiodic flows.
In Figure  \ref{meddie}c), for large $\tau=72$ days, the structure of $M$ in the interior of the eddy  becomes less and less smooth, meaning
 that in the range $(t-\tau, t+\tau)$ trajectories placed at the interior core have either concentrated there from the past or will disperse in the future. 
In fact,  the interior of the core is completely foliated by singular features associated either to stable or unstable manifolds of nearby hyperbolic trajectories.
The non-smoothness of $M$ at $t={\rm April}\,\, 17$ proposes  $2 \tau=144$ days as an upper limit for the time of
residence of particles in the inner core;  
particles perceive nearby hyperbolic regions after this period.  
The accuracy of the singular lines of $M$ representing  invariant manifolds is again  confirmed in Figure \ref{overlap}, where
computations of stable and unstable manifolds overlap those features. 
The foliated structure of $M$ is much richer than  that provided by the displayed manifolds computed directly. This is so
 because  the direct computation of  manifolds requires the location of {\it a priori} special hyperbolic trajectories (also called DHTs as explained in next section)  
 from which the manifold calculation starts.  The selection of DHTs may leave out many other DHTs in the neighbourhood while 
 $M$ exhibits all stable and unstable manifolds
 from all possible DHTs in the vicinity of the eddy, without the need for  identifying them a priori.
$M$ provides the complete {\it visible}
 foliation in the interval $(t-\tau, t+\tau)$ induced by the stable and unstable manifolds of all nearby hyperbolic trajectories.   

\begin{figure*}
a)\includegraphics[width=16cm]{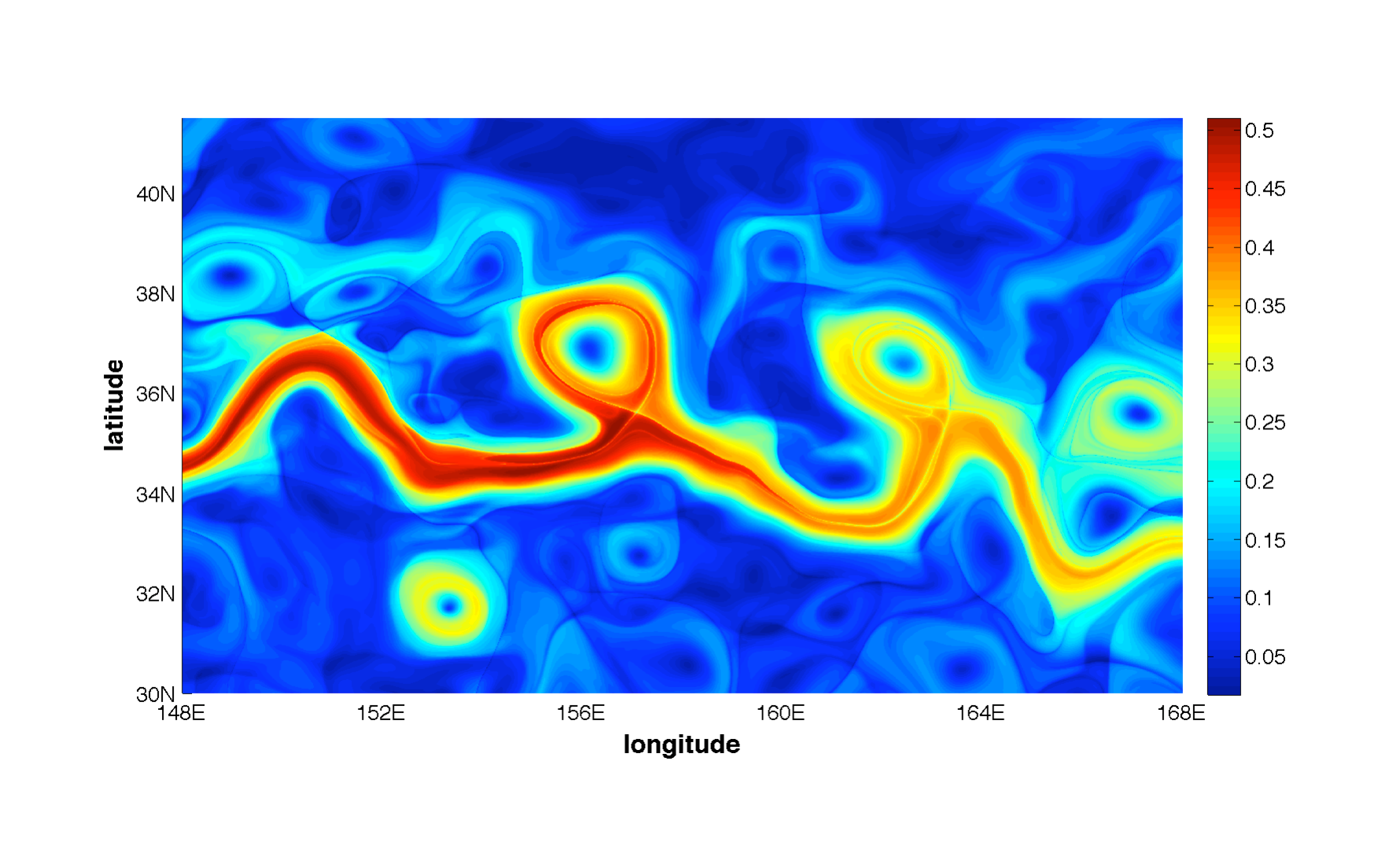}
b)\includegraphics[width=16cm]{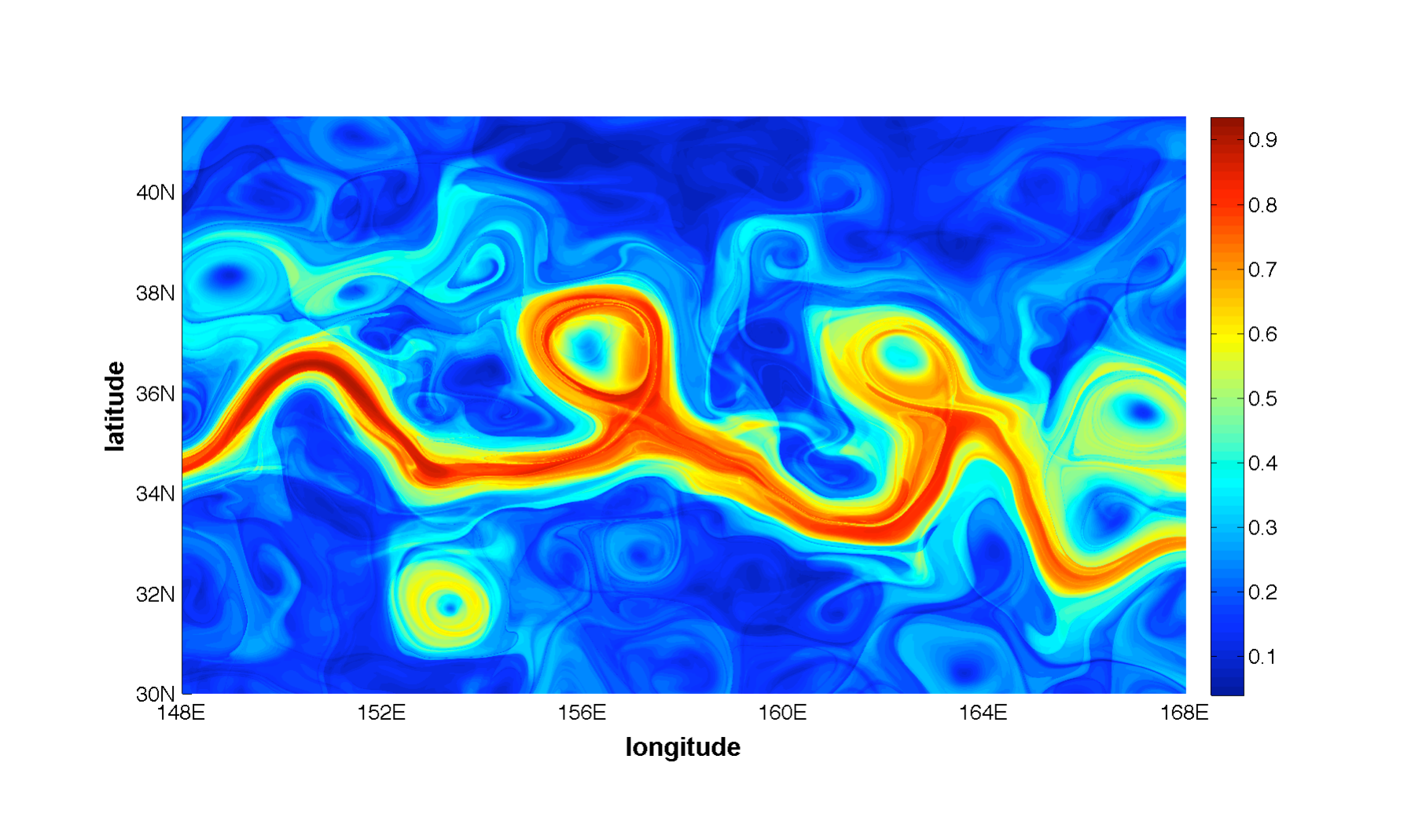}\\
\caption{\label{fig:M1}
Evaluation of the function $M$ over the Kuroshio current  between longitudes $148^o$E-$168^o$E and latitudes
 $30^o$N-$41.5^o$N on  May 2, 2003; a)  $\tau=15$ days; b)  $\tau=30$ days. (Figure taken from \cite{prl}).}
\end{figure*}

The evaluation of $M$ in large oceanic areas for long enough $\tau$, as shown in Figure \ref{fig:M1}, 
reveals recognisable phase portraits. The colour gradation of $M$ emphasises  lasting and stronger features versus 
the ones that are weaker and more transient.
  Largest $M$ values are in red while the lowest  are in blue. 
For instance in Figure \ref{fig:M1}a) the colours indicate  that
the strongest features  are a  central reddish  stream and the one red and  two yellow eddies. These are the most persistent patterns 
and because they remain for long periods of time it is possible to describe transport routes across them.
Other recognisable
bluish features such as the cat's eyes at the upper left  correspond to slow fluid motion. These features have  a rapidly 
changing topology, and    their lack of permanence  makes  it more difficult 
 to describe  transport across them, since transient structures are not well understood from the dynamical point of view (see  \cite{jpo,bra}). The function $M$ provides a global descriptor where different geometries  of exchange are visualised  in a straightforward manner.  Figure \ref{fig:M1}b) shows the output of $M$ at the same area, at larger  $\tau$ values. 
 A more complex structure near  the set of chaotic saddles is observed.   The increasing of complexity of  $M$  versus $\tau$ is  expected
 from the nature of $M$, since it is reflecting the history of initial conditions on open sets, and in highly chaotic systems this history is expected to be more tangled
 for longer time intervals.

Equation   (\ref{def:M}) proposes the integration along trajectories of  a bounded positive intrinsic geometrical or physical property. Imposing the integration of a positive quantity is consistent with
 the perspective that Lagrangian descriptors reveal the dynamical structure by accumulating quantities along trajectories. When trajectories separate following different paths
 the accumulated quantity differs, and sharp changes on the descriptor values should occur at the boundaries of regions separating these qualitatively distinct behaviors, thereby highlighting the position of  invariant 
  manifolds. 
 The accumulative perspective taken by Equation   (\ref{def:M}), although similar in its mathematical  expression, is different from the finite-time average velocities used in \cite{mezic2, mezic1}. In particular, these works consider the forward time integral of 
 the velocity components divided by the time interval:
 \begin{equation}
 \frac{1}{\tau} \int^{t^*+\tau}_{t^*} \! \!\!  v_x({\bf x},t)dt  
\end{equation}
 This averaging is reported to reveal a patchiness structure which is also
 connected to invariant manifolds.  In \cite{mezic1}, the authors note that for increasing averaging time a zero average velocity is obtained, and as a consequence  in this limit the spatial
structure in the patchiness plots is lost. 
As regards  the integration time limits  and their impact on the retrieved Lagrangian structure, the results by \cite{mezic1}
are the opposite of those  obtained  from the proposed  Lagrangian descriptors. We have reported the existence of a minimum time $\tau$ to converge to the Lagrangian structures, which is not reported by \cite{mezic2, mezic1},
and we have shown evidence  that beyond that $\tau$,  the longer  $\tau $ is, the better and more detailed are  the Lagrangian structures. 
The main reason for differences in the outputs  between both methods is  that the diagnosis by  \cite{mezic1}  does not force the integral of a positive quantity,  thereby allowing oscillations of the integrated quantity along trajectories, which produces non desired cancellations. Recently alternative methods which similarly to Lagrangian descriptors are based on measures along trajectories, have been  described in \cite{rypina}. These methods have been successfully applied to describe Lagrangian Coherent Structures in geophysical flows.

   \begin{figure*}
a)\includegraphics[width=6.5cm]{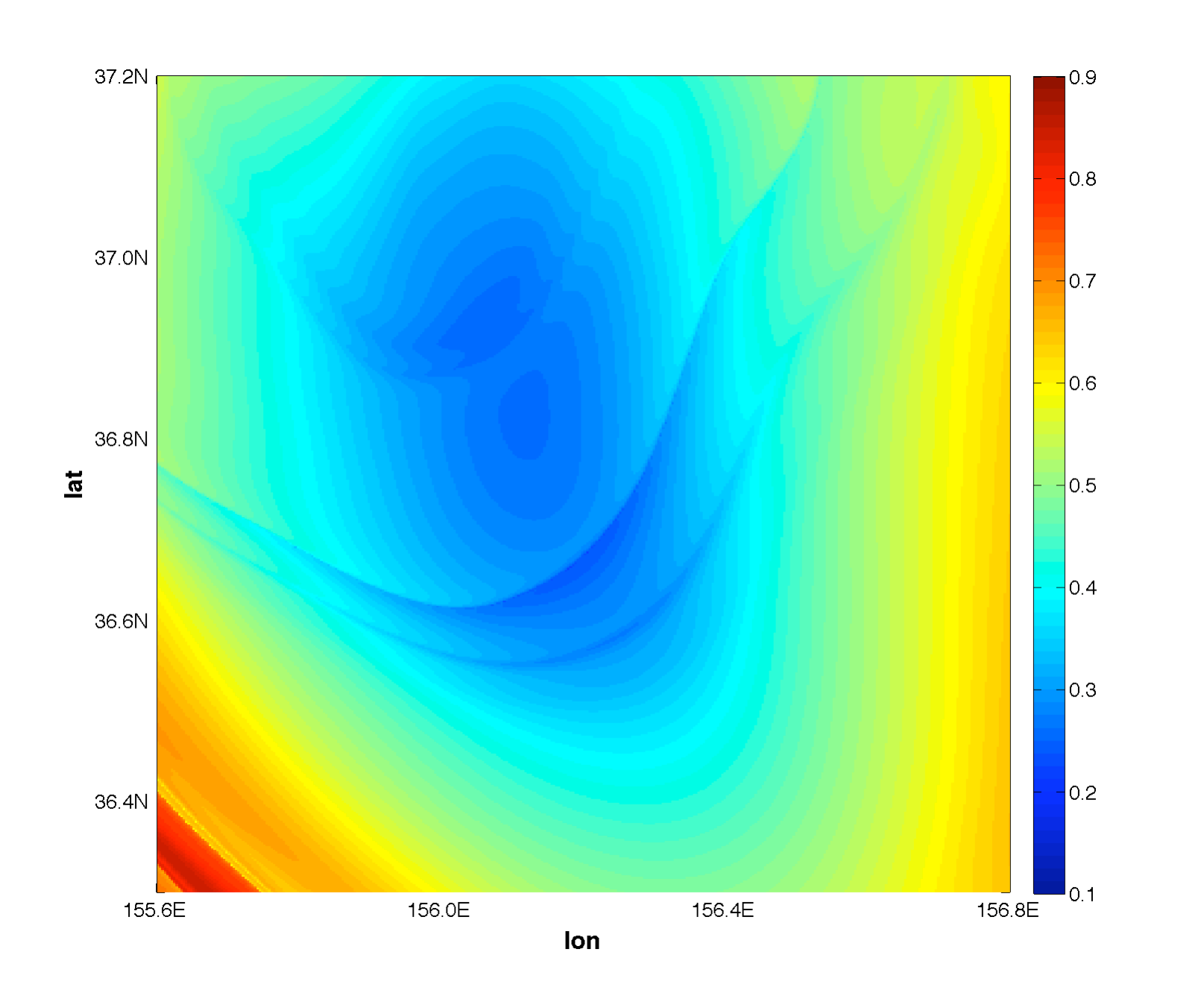}b)\includegraphics[width=6.5cm]{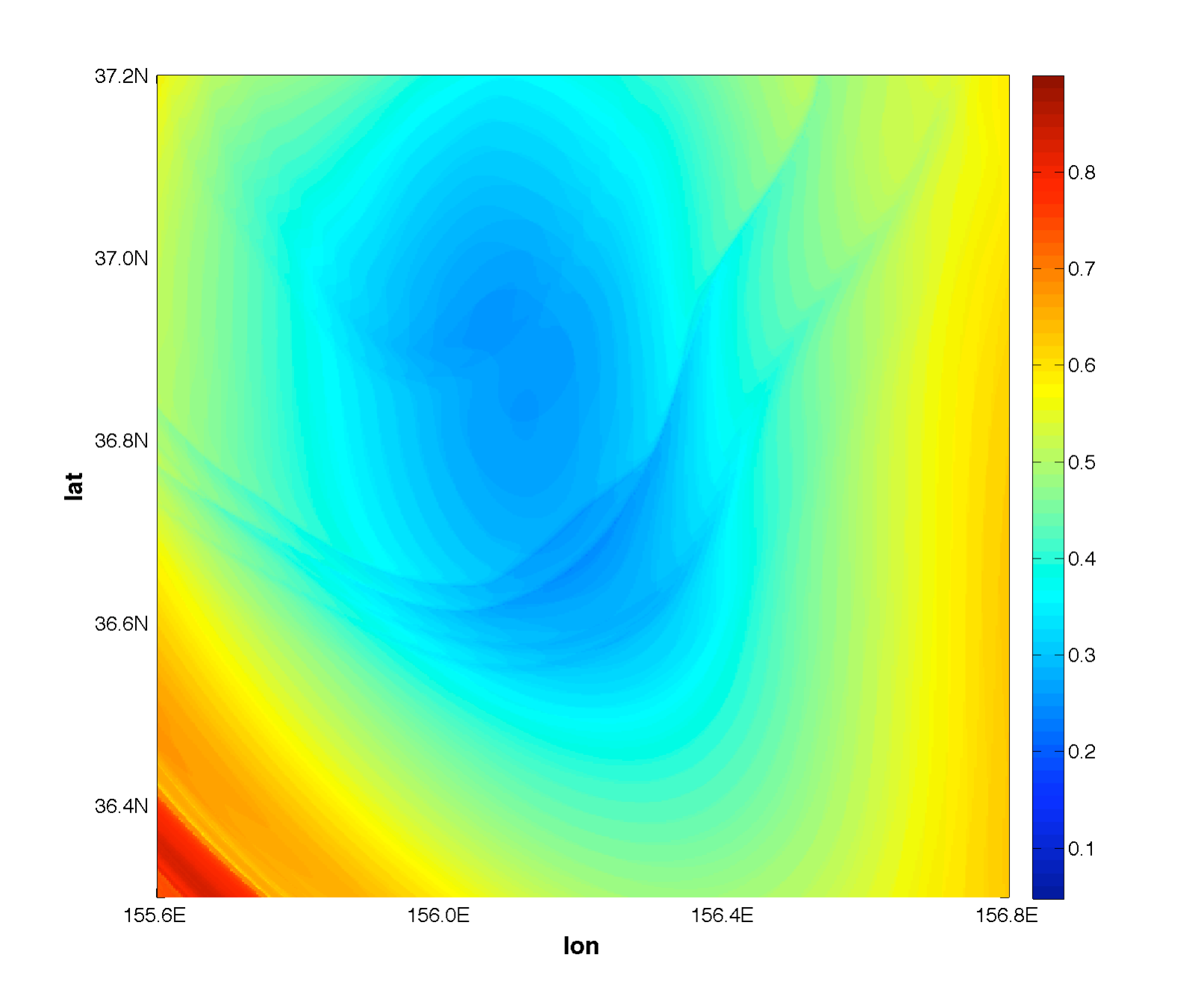}
\caption{\label{figlin}   Contour plot of the   function $M$ over the inner part of an eddy on  May 2, 2003 for $\tau=30$ days a) results with bicubic spatial interpolation; b) results with bilinear spatial interpolation.  }
\end{figure*}
 A question always under scrutiny is the robustness of the Lagrangian structures under errors.
 In the literature some results are found on this matter. For instance, \cite{ismael}  have studied the robustness
of the Lagrangian structures under deviations induced in the vector field by noise and dynamics of unsolved scales. They have confirmed the permanence of the FSLE features under these perturbations. 
 It is not our purpose to perform an analogue study on the function $M$. However  Figure \ref{figlin} presents some results in this regard. This figure estimates  the reliability of $M$  by computing it with different 
interpolation schemes: bi-linear and bi-cubic spatial interpolation.
 The displayed results are structures obtained at large $\tau$ in the inner part of an eddy. The bi-linear spatial interpolation preserves 
 the features obtained by the spatial bi-cubic interpolation, although it also adds some lines visible at the centre. Nevertheless   the global appearance of the output is preserved.

The global dynamical picture provided by the function $M$ enables us to foresee active transport routes over the ocean surface.   However, for describing 
detailed transport mechanisms  associated to the recognisable phase portraits,
 the intricate curves making up  manifolds must be accurately  computed over the ocean surface. Extracting these curves from the above embroiled  pictures
is a difficult and imprecise task, doomed to failure,  and for this reason we proceed   in a different  way, which is  explained in the following section.

\section{Distinguished trajectories and finite time invariant manifolds}

The role of $M$ in  transport description is based on its ability to cover the ocean surface with a geometrical
structure that resembles a patchwork of interconnected dynamical 
systems,  which indicates transport routes  to be described in further detail. This important capacity
 cannot be achieved by the tools described in this section, which only  
 provide details after the details themselves have been roughly identified {\it a priori}. Without this previous knowledge, the use of these  tools 
is less effective because they are too focused and  blind for distinguishing their own starting point.  On the other hand the detailed transport routes 
reported by the tools described in this section cannot be obtained just  by the use of Lagrangian descriptors.
The scenario displayed by $M$ in Figure \ref{fig:M1}a) shows a strong 
jet, visible in  the  intense reddish band, and two eddies  --interacting with the jet--  which are visualised by two  circles: one reddish situated towards the west side and the other yellowish to the east.  
For a detailed study of transport  in this area  we compute distinguished trajectories and manifolds.

\subsection{Distinguished trajectories }

The stable and unstable manifolds  of special hyperbolic trajectories such as  fixed points in autonomous dynamical systems
 or periodic trajectories in periodically time dependent  systems are the ones of  interest in our study. These  trajectories,  which act as organizing centres of the flow,
 do not have a natural extension for
 time-dependent aperiodic dynamical systems, in which
a generalization of these concepts is required. 
The definition of distinguished hyperbolic trajectories (DHTs) has succeeded to this end. Several definitions have been proposed, for instance see \cite{kayo,ju,chaos}.
In this article we follow the approach to these trajectories reported by \cite{chaos}, which is based on the Lagrangian descriptor given by the function $M$ in Eq. (\ref{def:Mgen}).  

\begin{figure*}
\begin{center}
a)\includegraphics[width=6cm]{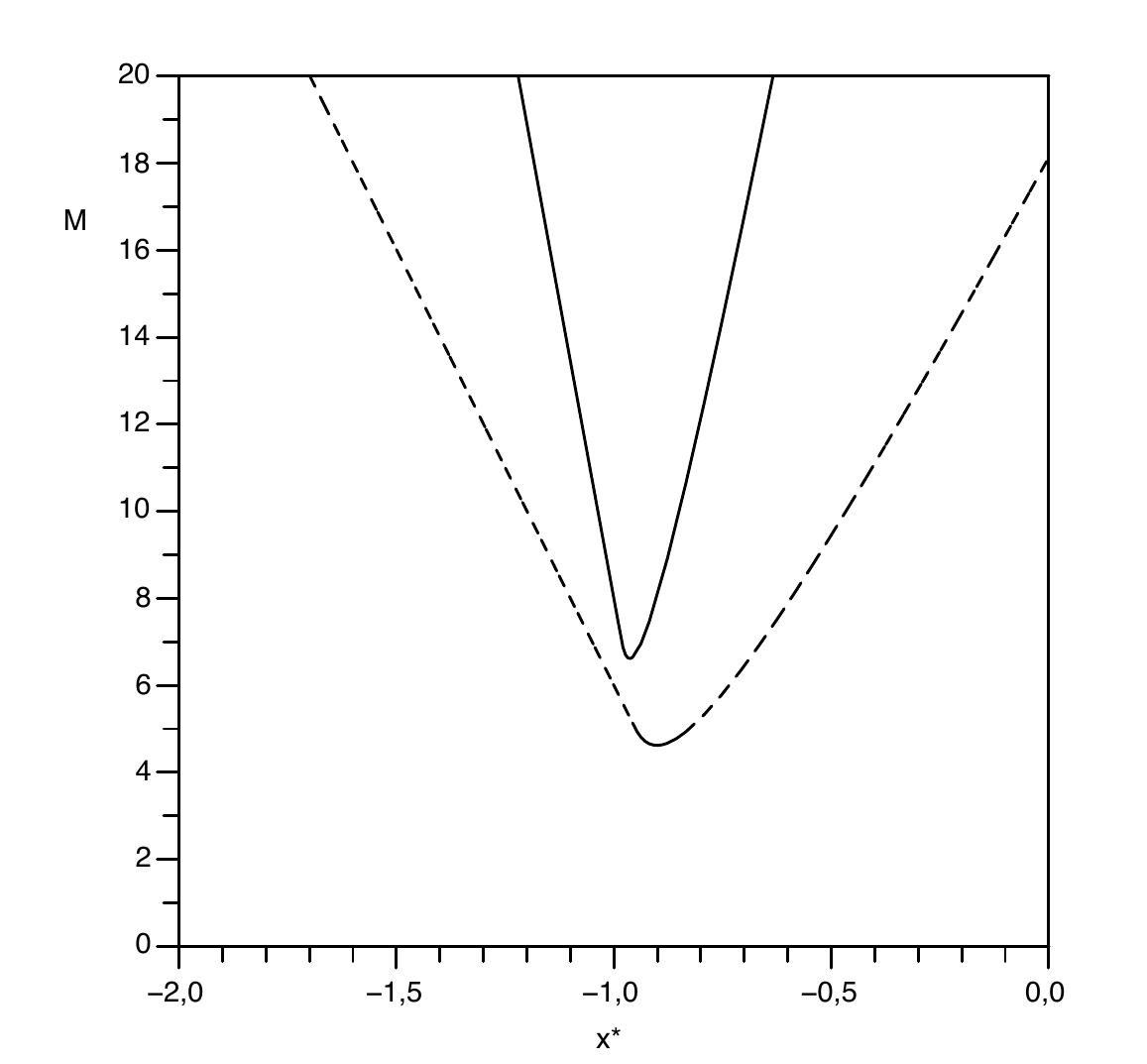}b)\includegraphics[width=6.9cm]{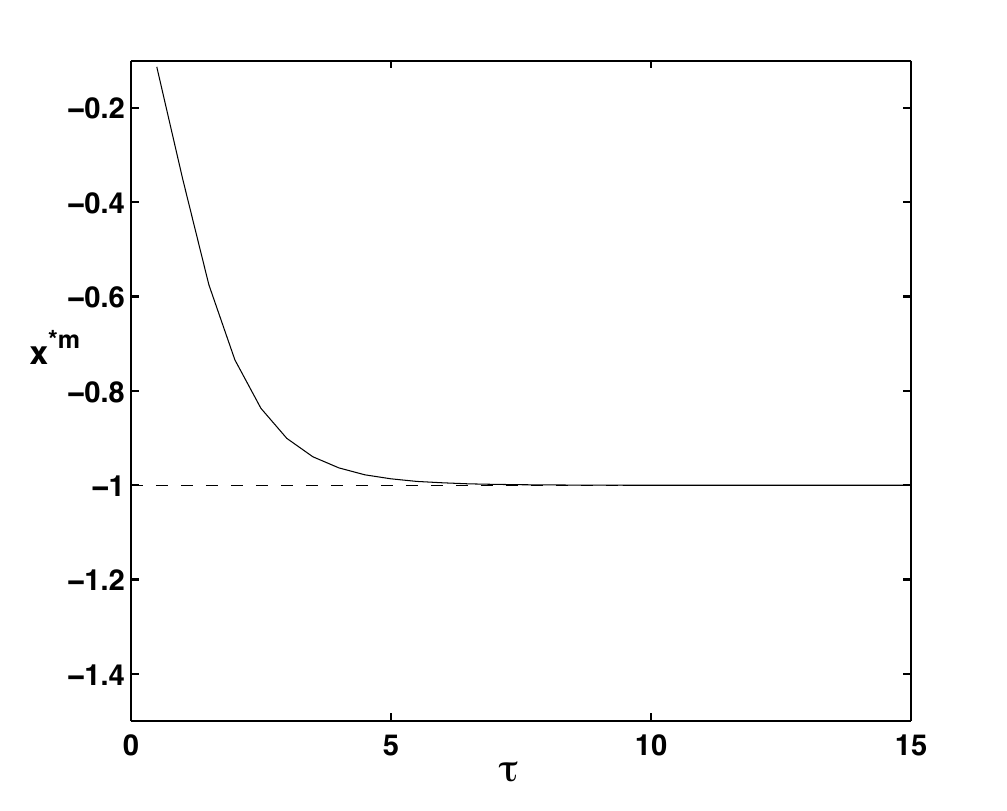}
\end{center}
\caption{ a) The function $M$ at $t=0$ for  $\tau=3$ (dashed line) $\tau=4$ (solid line); b) evolution of the coordinate ${\bf x}^{*m}$ at which  $M$ reaches a minimum  versus $\tau$ at $t=0$. (Figure taken from \cite{chaos}).}
\label{fig:dht1d}
\end{figure*}

The concept of DT generalizes the idea of fixed point for time-dependent dynamical systems. For instance for the 1D time-dependent linear system:
 \begin{eqnarray}
\frac{dx}{dt}=-x+t\label{1d}
\end{eqnarray}
the particular solution $x_p(t)=t-1$ is a generalized fixed point.  It is considered so, because the equation (\ref{1d}) by means of the Galilean transformation:
 \begin{eqnarray}
x'=x+vt \label{gtr}
\end{eqnarray}
is converted into  the autonomous system:
 \begin{eqnarray}
\frac{dx'}{dt}=-x'-1 \label{1daut}
\end{eqnarray}
which has a fixed point at $x'_p=-1$. The Galilean transformation (\ref{gtr}) applied to this fixed point transforms it back into the particular solution $x_p(t)=t-1$.
The intuitive geometrical idea behind our definition  for identifying $x_p$ as distinguished 
is to search for a  trajectory that "moves less" than others in a vicinity. But what does this mean?.  
For a given initial condition  ${\bf x^*}$ on an open set ${\mathcal B}$ at a given time $t^*$, ``move less" is satisfied by a trajectory that minimizes $M$ in Eq. (\ref{def:Mgen}).  This function
 measures the arc-length of the curve outlined on the phase space by the trajectory passing through $({\bf x}^*, t^*)$  from $t^*-\tau$ to $t^*+\tau$. In Figure \ref{fig:dht1d}a), $M$ is represented
for the system (\ref{1d})  at $t=0$ for $\tau=3,4$.  It is observed that $M$ reaches a minimum at different positions ${\bf x}^*$ for different $\tau$. However although this  fact may involve ambiguity in locating
the position for a DT, at large $\tau$ the position of the minimum converges towards what  is called the {\it limit coordinate}. Figure \ref{fig:dht1d}b) confirms this point.  There the position ${\bf x}^{*m}$ at which  $M$ reaches its minimum
is plotted,  which for increasing $\tau$ approaches the value $x=-1$.  This is exactly the passing point of the particular solution $x_p$ at $t=0$. In practice as noted by \cite{chaos} the convergence
to the limit coordinates cannot be examined in the limit $\tau \to \infty$ either because it is impracticable in a numerical implementation,
or because in the large limit errors accumulate,
or simply because the dynamical system is defined by a finite
time data set. For these reasons the convergence to the limit
coordinates is tested up to a finite $\tau$.  Formally this is expressed as follows: Let  us consider a practicable time interval $[T_i, T_f]$, let be ${\bf x}^{m*}_{t_l}(\tau)$ the coordinates at which the function $M$ 
reaches the minimum value at time $t_l$ in an open set  ${\mathcal B}$. Then to find the {\it limit coordinate} ${\bf x}^l$ at time $t_l$ we verify that there exists a $ \tau^l $ such that: $t_l-\tau^l >>T_i $, $t_l+\tau^l<<T_f$ and
$\forall \tau> \tau^l$  the following is satisfied: $||{\bf x}^{m*}_{t_l}(\tau)-{\bf x}^l(t_l) ||\leq \delta$ (where
$\tau $ keeps $t^l-\tau>T_i$ and $t_l+\tau<T_f$  and $\delta$ is a small positive constant). 
Here $|| \cdot ||$ represents the distance defined by
\[||{\bf a}-{\bf b} ||=\sqrt{\sum_{i=1}^{n} (a_i-b_i)^2} \,\,\,\, \,{ with} \,\,\,\,{\bf a}, {\bf b} \in \mathbb{R}^n.\]

By repeating the procedure at different times $t$, it is possible to obtain a {\it path of limit coordinates} which is denoted as ${\bf x}^l(t)$. The  {\it distinguished trajectory} $\gamma(t)$
is thus defined in a time interval $[t_0, t_N ]$ as that trajectory that is close enough (at a distance $\epsilon$) to a path of limit coordinates. According to \citep{chaos}  this is expressed formally as
follows:  A trajectory $\gamma(t)$ is said to be Distinguished with accuracy $\epsilon$ ($\,0\leq \epsilon$ )
in a time interval $[t_0, t_N]$ if there exists  a continuous path  of {\it limit coordinates} $(t^l,{\bf x}^l)$  where
$t^l \in [t_0, t_N]$, such that,
\begin{eqnarray}
||\gamma(t^l)-{\bf x}^l(t^l)|| \leq \epsilon, \,\, \forall t^l \in [t_0, t_N]
\end{eqnarray}\label{def:dtdef}
In this definition $\epsilon$ is a small positive constant within the numerical accuracy we can reach. Further examples of trajectories characterized as 
Distinguished are discussed  in the work by \cite{chaos} in two and three dimensions. 
\begin{figure*}[t]
\vspace*{2mm}
\begin{center}
a)\includegraphics[width=6cm]{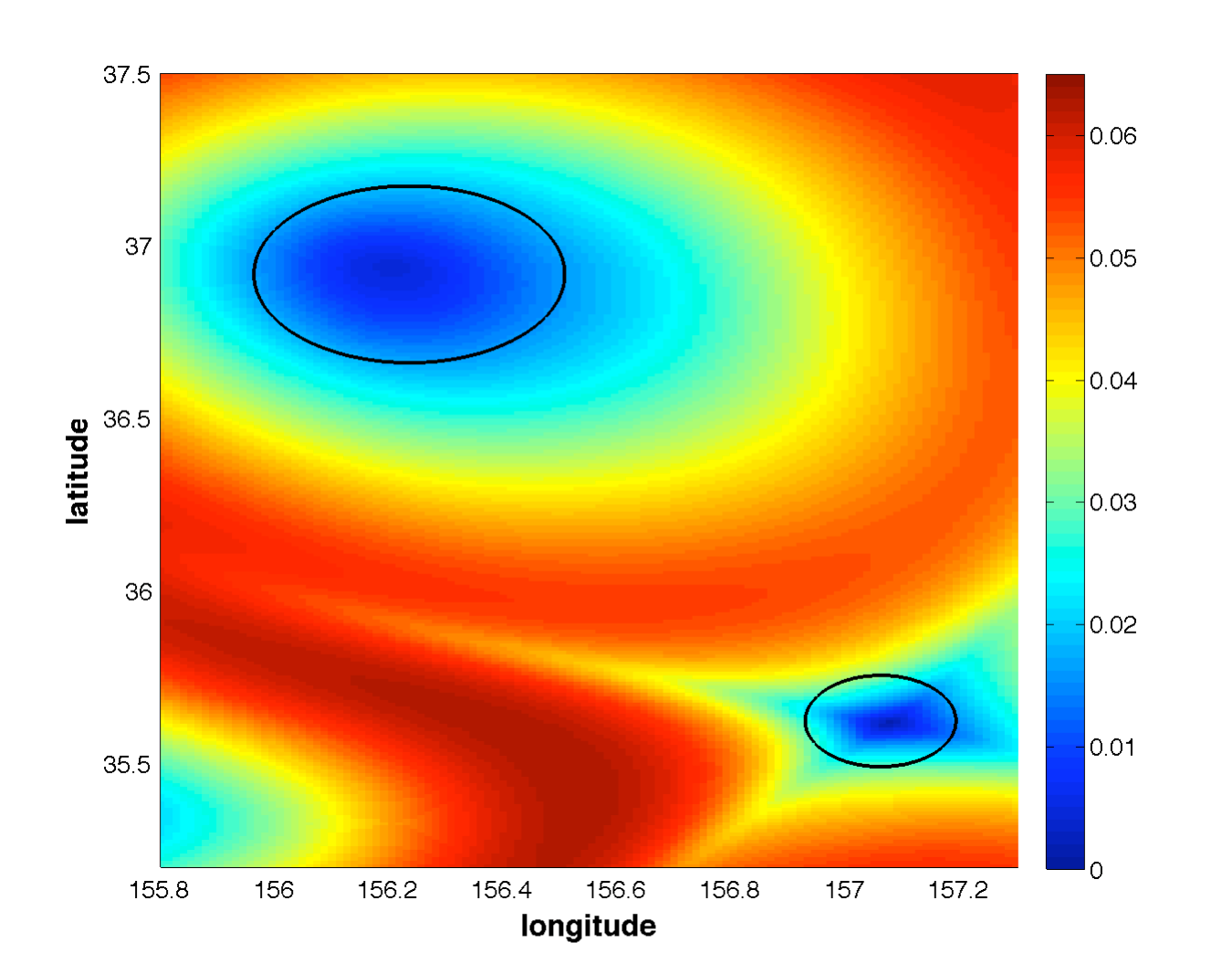}b)\includegraphics[width=6cm]{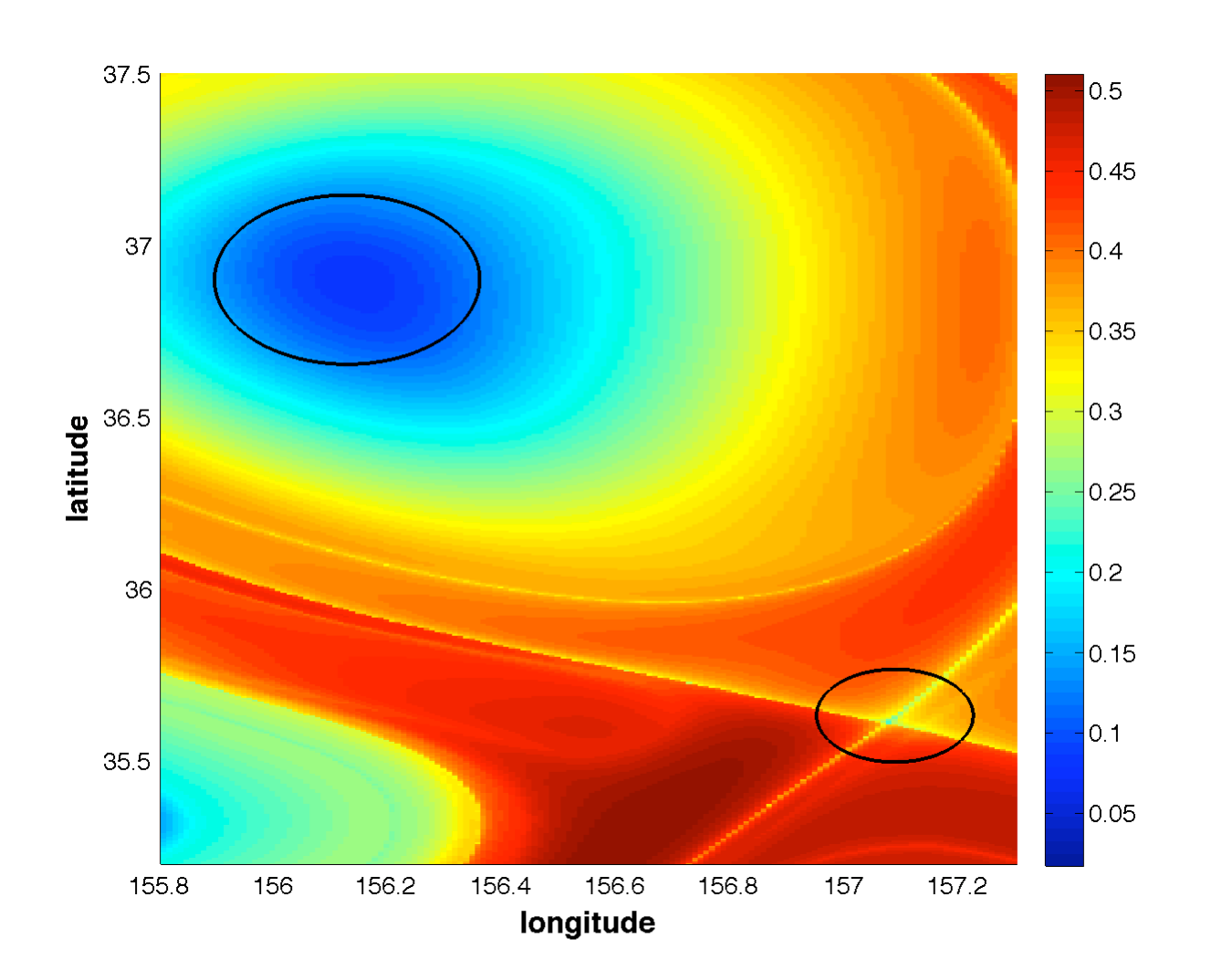}
\end{center}
\caption{ Contour plots of the function $M$ on  May 2, 2003 in the nearby of two positions which are candidates to be DT. a) $\tau=2$ days; b) $\tau=15$ days.}
\label{fig:M0cbis}
\end{figure*}

 \begin{figure}
\includegraphics[width=8.5cm]{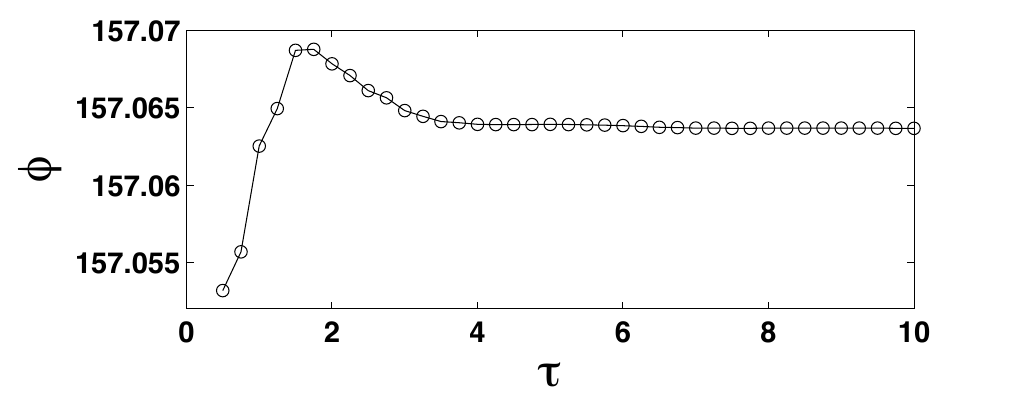}\\
\hspace*{0.5cm}\includegraphics[width=8.2cm]{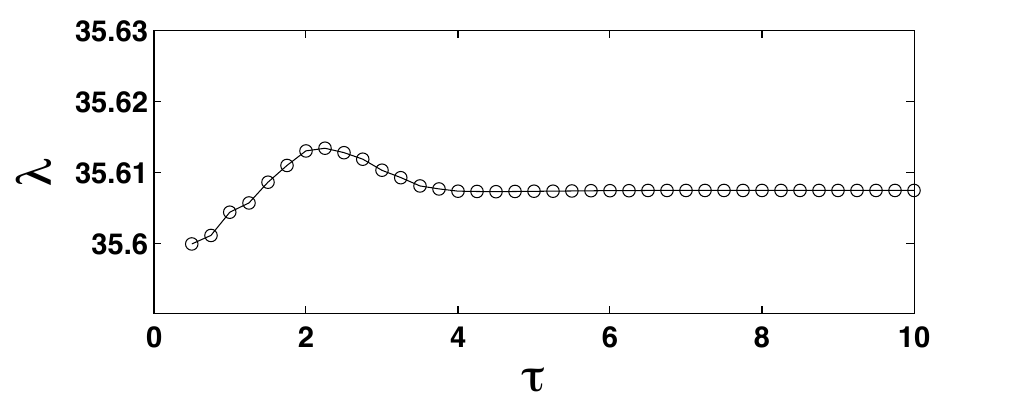}
\caption{\label{dhtlonlat} Evolution of the longitude and latitude position of the hyperbolic minimum of  the  function $M$ on  May 2, 2003 versus $\tau$ (in days).}
\end{figure}
\begin{figure*}[t]
\vspace*{2mm}
\begin{center}
\includegraphics[width=12cm]{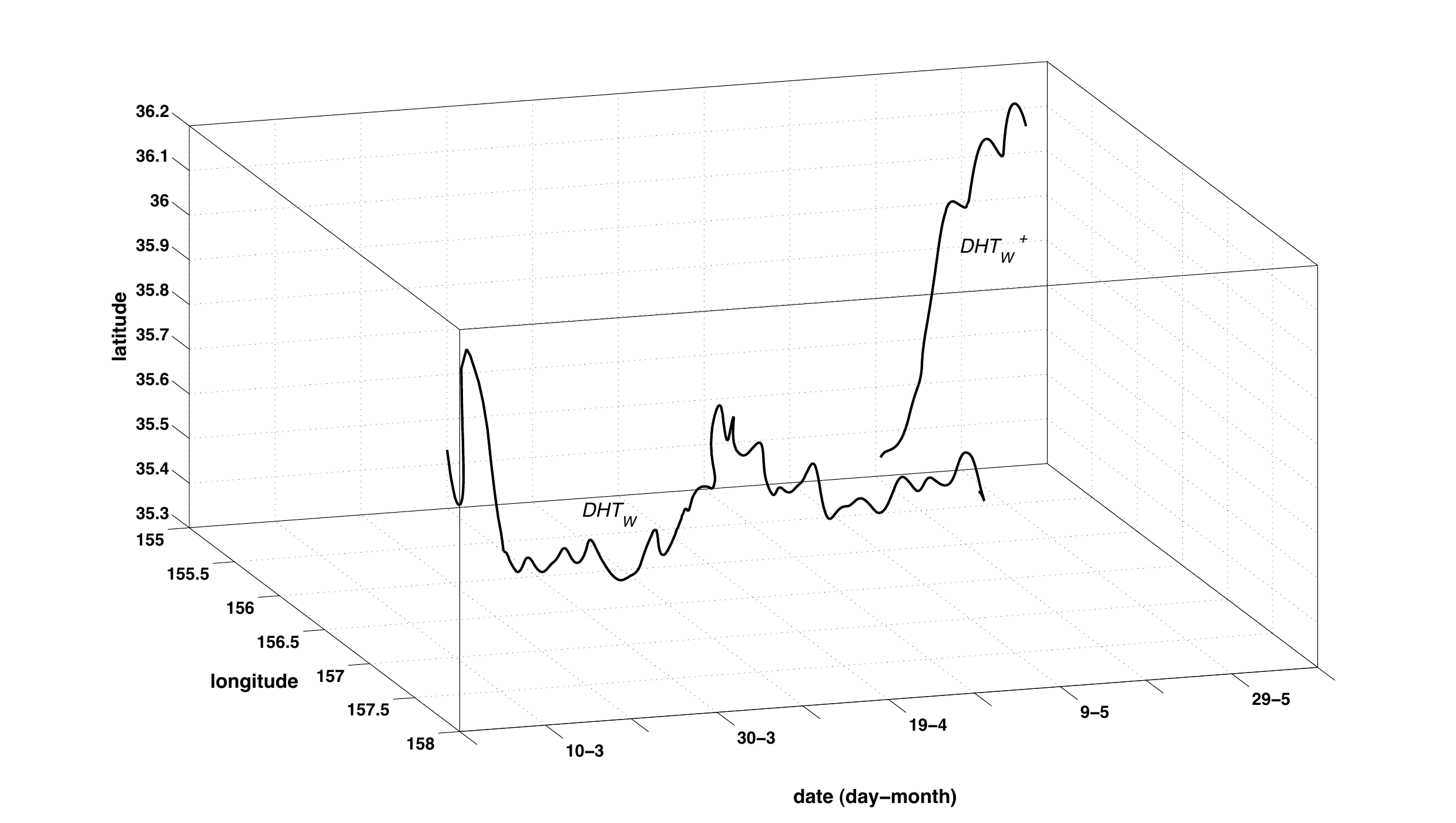}
\end{center}
\caption{ Path of limit coordinates for DHT$_W$ and DHT$^+_W$.}
\label{dhtl2}
\end{figure*}

Next we illustrate  how to  identify DT in our 2D data set.
Figure
 \ref{fig:M0cbis}a) shows a  contour plot of $M$  on $t^*=$May 2, 2003 for $\tau=2$ days in the neighbourhood of the western eddy. Two circles
surround the two minima of this open set. These minima correspond to initial conditions whose trajectories  outline curves  shorter  on the phase space
 than those in their vicinity.  Figure
 \ref{fig:M0cbis}b) shows the same contour plot  of $M$, but at $\tau=15$ days. A comparison with figure  \ref{fig:M0cbis}a) reveals several differences. 
  The neighbourhood of the minimum in the lower circle     of Fig \ref{fig:M0cbis}b) presents a crossed-line  structure,
 that has been linked to manifolds. In the interior of this structure there exists a minimum whose position does not coincide
 with that obtained at $\tau=2$ days. Figure \ref{dhtlonlat} shows the evolution of the longitude and latitude position of the minimum with $\tau$ converging to a limit coordinate.
 In figure  \ref{fig:M0cbis}b), the minimum in the lower
circle has reached the position of the   {\it limit coordinate} within the accuracy $\epsilon$ available with our numerical schemes.
It is possible to track  in a set of discrete times $t^l$ the path $(t^l,{\bf x}^l)$ described by this limit coordinate in a time interval. 
The path    is displayed in Figure \ref{dhtl2}. 
 In the vicinity of the path displayed in Fig \ref{dhtl2} at a distance $\epsilon$    is found DHT$_{W}$, a trajectory
that remains distinguished from March 5 to May 11, 2003.  Figure \ref{dhtl2} also represents the coordinates of a second trajectory labelled as DHT$^+_{W}$, in an almost complementary period of time, between May 10 and June 1, 2003. Trajectories  DHT$_{W}$ and  DHT$^+_{W}$ were first characterised in this data set 
 by \cite{nlpg2}. 
 By construction, a Distinguished Trajectory defined in this way is a property held by some 
 trajectories in finite time intervals. Alternative definitions such as those provided  in \cite{kayo,ju} do not address this possibility.

\begin{figure}
\begin{center}
\includegraphics[width=4cm]{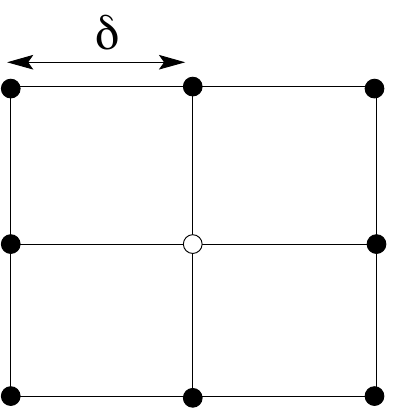}
\end{center}
\caption{\label{grid} The grid used to find  the minimum position with precision $\delta$. The white central dot indicates the position where the minimum is due.}
\end{figure}

The  ideas described above are itemised in the algorithm that computes DT,  and is fully described in  \cite{chaos}. We give a brief account of it next. 
It  starts by estimating an approximate position ${\bf x}^{*m}$ for the minimum   of $M$ at low $\tau$  
in a specific area  at a given time $t^*$. Its coordinates are refined 
up to a precision $\delta$ by considering  a grid such as that depicted in Figure \ref{grid}, which has its centre positioned at  ${\bf x}^{*m}$. $M$ is evaluated in the nodes of the grid and if the lowest  $M$-value is not taken  at the centre, 
 but in a peripheral node the grid  displaces its centre at this position of the minimum, which provides a better approach for ${\bf x}^{*m}$.  $M$ is then reevaluated in the nodes of the new positioned grid, and
 if the minimum is found to be  at the central node, the search stops. This method is used to follow 
 the position of the minimum at iteratively increasing $\tau$:  $\tau_{k}=\tau_{k-1}+\Delta \tau$, where $\Delta \tau$ is a small quantity. The procedure 
 stops when the position of the minimum ${\bf x}^{*m}$ does not change for further $\tau$ increments. 
   At  the next time  $t^*+\Delta t$,  ${\bf x}^{*m}$  is time evolved with the equations of motion, and the iterative search 
   described above starts from this point. 

\begin{figure}
\includegraphics[width=10cm]{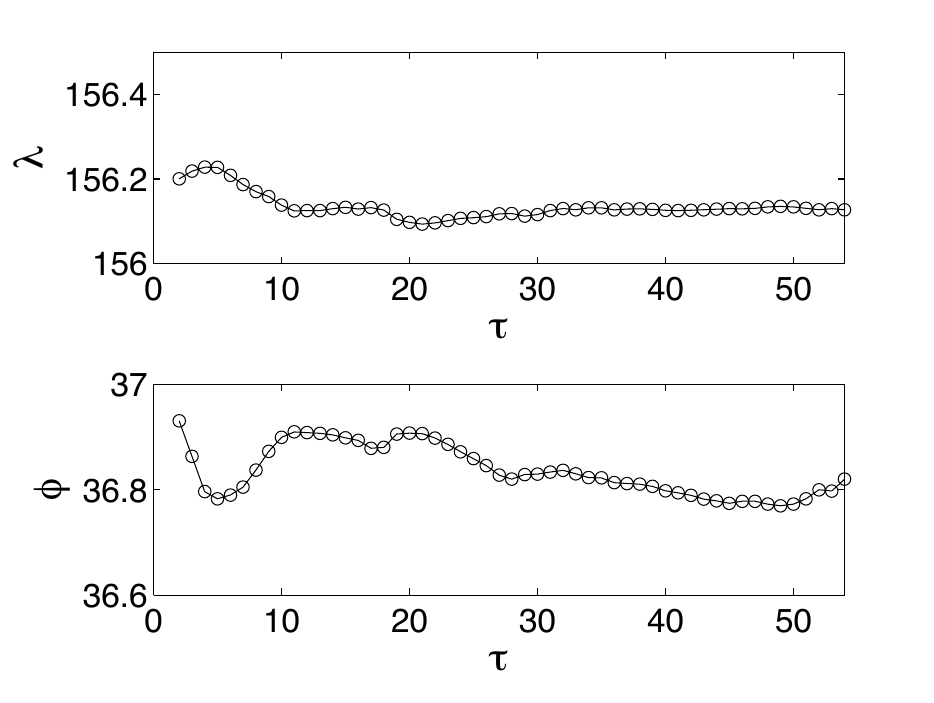}
\caption{\label{det} Evolution of the longitude and latitude position of the elliptic minimum of  the  function $M$ on  May 2, 2003 versus $\tau$ (in days).}
\end{figure}

The  minimum situated in the interior of the upper circles in Figure  \ref{fig:M0cbis} presents a structure that evolves with $\tau$ quite differently
 to what is found in the lower circles. It remains  rather flat and circular and does not evolve towards the crossed line structure typical of  DT with hyperbolic stability (DHT). As discussed in  \citep{chaos} these 
 patterns are typical  of a DT with elliptic stability (DET).
Figure  \ref{det} shows the evolution of the coordinates of this minimum versus $\tau$. In this case, a DET
is not properly identified, because  contrary 
to what is found for hyperbolic cases,  a limit coordinate is not reached.
DETs are not easily found  in highly aperiodic flows. A previous attempt has been discussed in \cite{chaos} for a different data set, and
a failure to satisfy the definition is reported. Successful examples of DET are however reported 
for time periodic dynamical systems  (see \cite{chaos} for full details).
Although this elliptic minimum is not related to a special trajectory, it   still locates a  coherent structure related to an oceanic eddy. 
As reported in the previous section, 
particle confinement on this  area  persists  in a time interval $[t^*-\tau, t^*+\tau]$ provided that $\tau$ is below the limit at which the foliation induced by the stable and unstable
manifolds of nearby hyperbolic trajectories penetrates the inner core. Precisely  the fact that these eddy-like structures  eventually perceive  nearby hyperbolic trajectories, would justify  the absence of DETs in their interior.

\begin{figure*}[t]
\vspace*{2mm}
\begin{center}
\includegraphics[width=12cm]{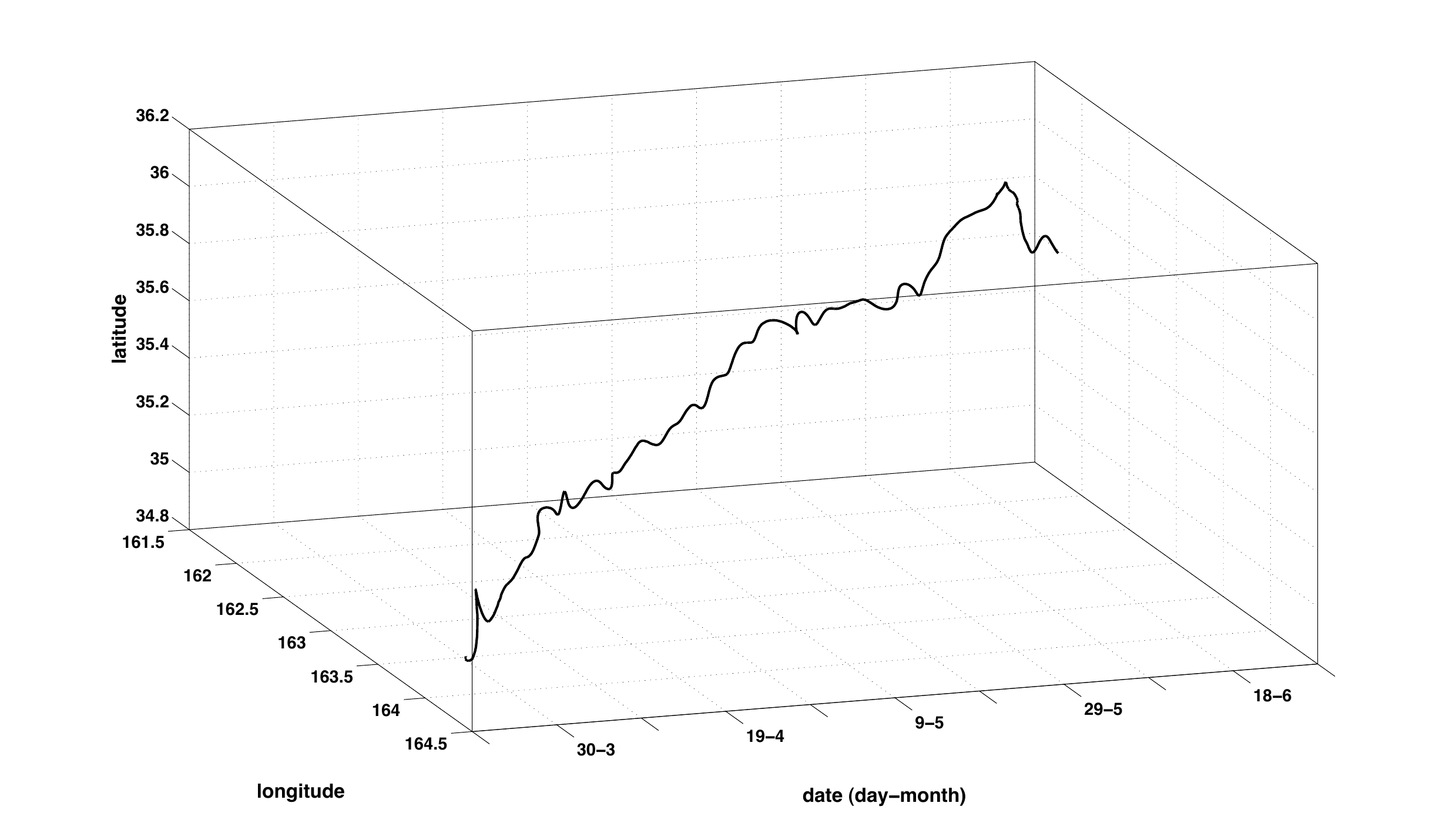}
\end{center}
\caption{ Path of limit coordinates for DHT$_E$.}
\label{dhtl3}
\end{figure*}

  In the scenario shown in by Figure \ref{fig:M1}a), at the east bound the jet    interacts with the yellowish eddy to form a crossed line structure  which is identified as an
eastern  DHT. The path of limit coordinates near  this DHT$_E$ is represented in Figure  \ref{dhtl3}.  It stays as distinguished  between March 25 and June 24, 2003 (see also \cite{nlpg2}). 

 \subsection{Finite time invariant manifolds}
 
 Invariant manifolds are mathematical objects classically defined for infinite time intervals. 
The unstable (stable) manifold of a hyperbolic  fixed point or periodic trajectory 
is formed by the set of trajectories that in minus (plus) infinity time approach these special trajectories. 
In geophysical contexts this definition is not
realizable, because on the one hand only  finite time aperiodic data sets are possible and on the other hand the reference trajectories, the DHTs,  
typically hold the distinguished property in finite time intervals. However, 
a detailed description of  Lagrangian transport requires a direct computation of 
 the stable and unstable manifolds of the selected DHTs. \cite{bra2} have recently proposed a novel algorithm to compute invariant manifolds in 
 3D non-autonomous dynamical systems. Nevertheless  our next presentation is focused on the illustration of this procedure in 2D flows as corresponds to the selected data set.
 \cite{nlpg,jpo,nlpg2,bra} have computed 
 stable and unstable manifolds of DHTs   for  2D highly aperiodic data sets  by   using the method  proposed in \citep{mani}. Based on  ideas and techniques  
 of contour advection \citep{dr,amb}, the
 algorithm computes manifolds  
 as  curves advected by the velocity field, which  at the beginning of the procedure are small segments 
 aligned with the stable and unstable subspaces of the DHT.
 The use of these small segments
in the starting  step is the way to build {\it a finite-time version of  the asymptotic property} of manifolds. 
Hence in our computations the finite-time unstable manifold at a time $t^*$ is made of trajectories that at time $t_0$, $t_0<t^*$ were at a small segment aligned with the unstable subspace of the DHT.
Similarly the finite-time stable manifold at a time $t^*$ is made of trajectories that at a time $t_N$, $t_N>t^*$ are in a small segment aligned with the stable subspace of the DHT.
 Localising thus a DHT and its stable and unstable directions at the starting time constitutes the first step. The way in which the stable and unstable subspaces are identified
 is closely related to the way in which DHTs are computed. For instance algorithms for DHTs described in \cite{kayo, ju} provide them directly as an output, and this is
 the start-up for the manifolds computed in  \cite{nlpg,jpo,bra}. The algorithm  for DHTs reported in \cite{chaos}, which is the one followed
 in this work,  does not provide these subspaces, but we note that stable and unstable subspaces are supplied by the crossed lines recognised in the contour plots of
 the function $M$  near the DHT. These lines, as reported in \cite{prl,nlpg2}, are advected by the flow and constitute a close-up of the manifold near the DHT.
Segments  within
the  stable and unstable subspaces of the DHT
are respectively evolved backward and forward in time to obtain the fully nonlinear stable and unstable manifolds \citep{nlpg2}. 
 We focus on describing the details for
obtaining  the unstable manifold, noting that the stable manifold is obtained in a completely analogous way by inverting the time direction. The unstable
manifold is represented at time $t_0$ by a set of points on the unstable subspace.  The manifold is computed in a discrete set of time increments $t_k$ for $k=0...N$,
in which it is represented by a well chosen set of points. We explain how to determine these points at every time  $t_k$.  
The procedure  starts by considering the points on the initial segment which are evolved in time from $t_0$ to $t_1$. As they evolve they may grow apart, giving rise to unacceptable large gaps 
between adjacent points on the manifold. The criterion for unacceptable gaps is given by a quantity $\sigma_j$ which is defined at each point ${\bf x}_j$
as the product of the distance $d_j$ between adjacent nodes ${\bf x}_j$ and  ${\bf x}_{j+1}$ times the density $\rho_j$, {\it i.e},  $\sigma_j = d_j \rho_j$. If $\sigma_j>1$
 the gap between nodes is unacceptable. The density of points along the computed manifold is measured by  $\rho_j$, for which 
 several expressions are proposed  \citep{dr, amb}. We consider it defined as in Eq. (\ref{eq:rho}) in the Appendix B.
When a gap between nodes at time $t_1$ is too large according to the criterion just defined, it is filled by inserting a point at $t_0$ between 
the same nodes using an appropriate interpolation technique. At this stage the interpolation could be  simple because the curve at $t_0$ is a straight line. However most refined interpolation techniques are required
when this procedure is applied to evolve the manifold from $t_k$ to $t_{k+1}$  for $k>0$, since manifold becomes more and more intricate.
The most successful  interpolation scheme of those used in \cite{mani,physrep} is due to \cite{dr}. This method  represents the
curve between points ${\mathbf x}_j$ and ${\mathbf x}_{j+1}$ as the polynomial given by Eq. (\ref{eq:drint}) in the Appendix B.
The criterion is verified for each pair of adjacent points making up the manifold at $t_1$ and the procedure is iterated  until there are no gaps exceeding the tolerance.
Once the gap size acceptability condition is satisfied at $t_1$ we use the point redistribution algorithm described
in \citep{dr} in an attempt to remove points from less computationally demanding parts of the manifold (see \cite{nlpg}). This algorithm works as we
describe in the Appendix B. The complete procedure to evolve the unstable manifold from $t_0$ to $t_1$ is repeated for successive times  $t_{k-1}$, $t_{k}$ until the end time $t_N$ is reached.
Stable manifolds are obtained in a similar way, but the computation is started at time $t_N$. 

\begin{figure}
\includegraphics[width=9cm]{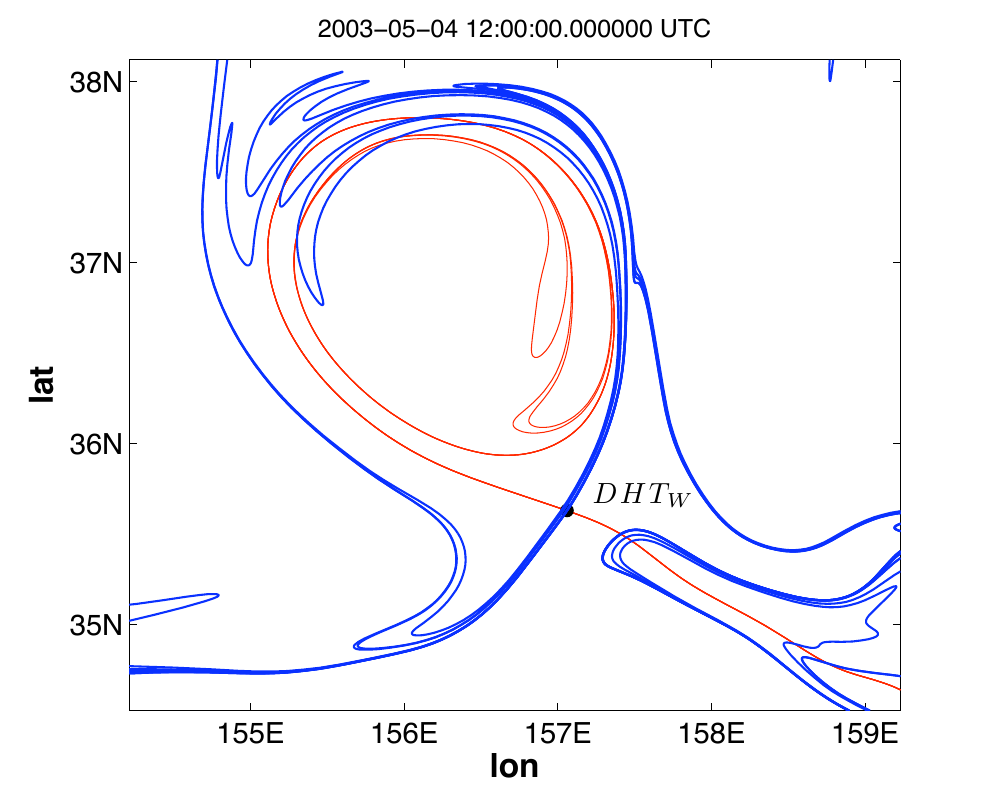}
\caption{\label{mani} Stable (blue) and unstable (red) manifolds of  DHT$_W$ on the 4th of May 2003. (Figure taken from \cite{nlpg2}).}
\end{figure}

Examples of manifolds computed with this method are shown in Figures \ref{overlap}  and \ref{mani} for the western DHT$_W$.
Manifolds computed in this way become very long and intricate  curves and from them   transport is described in great detail as discussed in the next section. 
Almost  every distinguishable line in Figure \ref{mani}
contains numerous foldings of each manifold, thus confirming how intricate they may be. Other approaches such as  FTLE or FSLE compute  manifolds at a given time  as ridges of a scalar field, thereby
 providing pieces of curves that are approximately material curves.  However, in these approaches links between pieces of curves are  difficult to establish as they  fade away and this is a disadvantage compared with  the direct computation of manifolds, which provides long complex linked curves due to the asymptotic condition imposed in their computation. 

As noted in the previous subsection, DHT$_W$ is characterized as Distinguished in a finite-time interval:  from March 5, to May 11 2003. A question  then to be addressed is,  what happens to the unstable manifold in Figure \ref{mani} beyond May 11, once DHT$_W$ losses its distinguishing property? It is observed that the manifold computation beyond this time may continue,  because  even if the reference trajectory on it is lost, the computation  still provides a material surface  advected by the flow, and second this advected 
object  is still  asymptotic to  DHT$_W$ in the finite-time sense introduced above. A similar argument can be made for the stable manifold in times prior to March 5. 
 DHTs and their stable and unstable subspaces are the starting step  of the  algorithm for direct computation of manifolds. However,  as reported in \cite{nlpg,physrep} they are not required by the algorithm beyond this point.  Nevertheless it  is useful  to have
  the full track of the DHT  for transport description purposes, because it marks a reference point which separates the manifold  into two branches. Section 5 illustrates  the application of this division.
  
\subsection{Frame invariance}

In this section we discuss the issue of `frame invariance''. To begin with, it is important to understand  what is meant by this phrase in the context of our work. There are two main issues. One is how the Lagrangian tools, such as those based on $M$,  perform in different coordinate systems. The other is how stability and geometrical features of the flow  transform under coordinate transformations.  
It is expected that under coordinate transformations the results obtained from the $M$ function will transform  according to the manner in which   the type of invariant objects  that the $M$ function is expected to recover transform.  However, we  note that in general these invariant objects are not preserved  under arbitrary coordinate transformations, as we will illustrate in this section.  Three examples will provide evidence of  these issues below. 

The function $M$ is used for two different purposes. One is discovering and visualising the global dynamics of a time-dependent velocity field  $M$ realises manifolds at positions at 
which abrupt changes in $M$ occur.  If the coordinates of a dynamical system are transformed to a 
rotating frame or to a frame moving with a constant velocity (i.e. a Galilean transformation of the coordinates) the
 manifolds will  transform to manifolds in the new coordinate system under the same transformation of coordinates. Of course
 the values of $M$ at specific points of space will certainly change with the reference frame,  but the edges at which $M$  changes abruptly -- which are the features containing the Lagrangian information-- are transformed with the change of coordinates in the same manner in which the  manifolds themselves are transformed. This is expected to be the case since the heuristic argument introduced to justify why $M$ detects manifolds is independent of a particular coordinate frame--manifolds play the role of dividing 
the phase space into regions corresponding to particles with qualitatively different dynamical fates and this is the case for any reference frame.

We verify this  argument for the periodically forced Duffing equation under the rotation:
 \begin{eqnarray}
R(t)=\left(\begin{array}{cc}\cos \omega t & -\sin \omega t \\
\sin \omega t  & \cos \omega t
 \end{array}\right).
\end{eqnarray}
\noindent
In the rotating frame  this equation takes the form:
  \begin{eqnarray}
\left(\begin{array}{c}\dot{\eta_1}\\ \dot{\eta_2} \end{array}\right) &=& \left(\begin{array}{cc}\sin 2 \omega t & \cos 2\omega t +\omega\\
\cos 2\omega t -\omega & -\sin 2 \omega t
 \end{array}\right)\left(\begin{array}{c}\eta_1\\ \eta_2 \end{array}\right) \nonumber\\
&+&( \varepsilon \sin t - [ \cos \omega t 
\eta_1 - \sin \omega t \eta_2 ]^3 )\left(\begin{array}{c}\sin \omega t \\ \cos \omega t \end{array}\right).
\end{eqnarray}

\noindent
The Duffing equation in the  non-rotating system:
 \begin{eqnarray}
\dot{x_1}&=&x_2\\ 
\dot{x_2} &=& x_1-x_1^3+\epsilon \sin(t)
 \end{eqnarray}
possesses a distinguished hyperbolic trajectory (DHT). This DHT can be computed as a perturbation expansion in $\epsilon$ about the hyperbolic fixed point in the $\epsilon =0$ case: 

\begin{equation}\label{eq:dhtduffing}
{\bf x}_{DHT}(t)=-\frac{\varepsilon}{2} \binom{\sin t}{\cos t}-\frac{\varepsilon^3}{40}
\binom{2\sin^3 t+\frac{3}{2}\sin t\cos^2 t}{\frac{3}{2}\cos^3 t+3\sin^2t\cos t}
+\mathcal{O}(\varepsilon^5).
\end{equation}

\noindent
The DHT  in the rotating frame ${\boldsymbol{\eta}}_{DHT}$ is  obtained by transforming the DHT in the non-rotating frame with the coordinate transformation:
\begin{equation}
{\boldsymbol{\eta}}_{DHT} (t) = R(t)^{-1} {\bf x}_{DHT} (t). \label{dhtrduff}
\end{equation}

\begin{figure*}
a)\includegraphics[width=6.2cm]{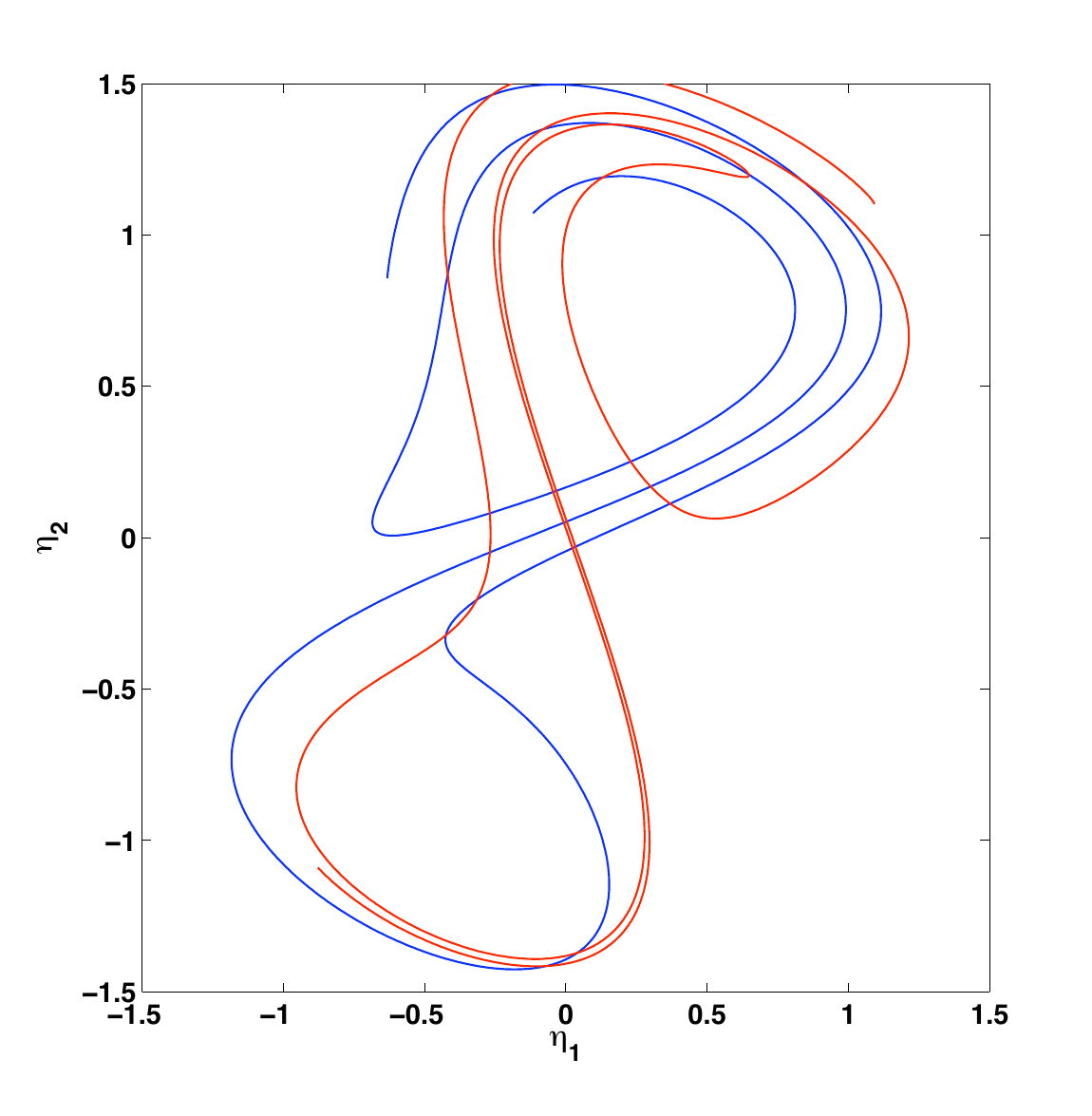}b)\includegraphics[width=7cm]{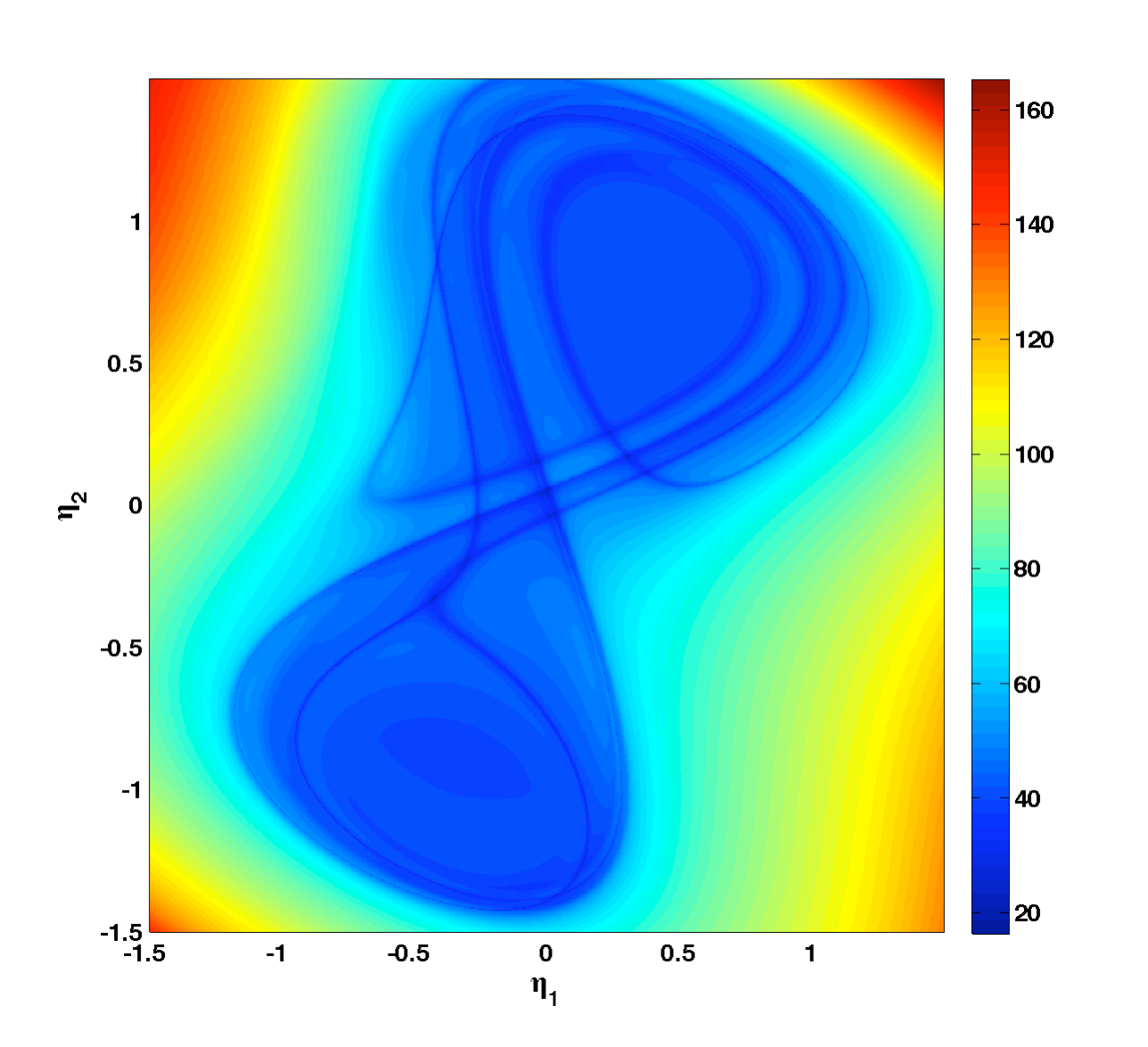}
\caption{\label{figrotduff} a) Invariant manifolds for the rotating Duffing equation for $\omega=2$
and $t=1$; b) contour plot of descriptor $M$ for  $\mathcal{F}=|{\bf v}|$ with $\tau=10$  at the same $\omega$ and $t$ values.}
\end{figure*}

\noindent
Stable and unstable
 manifolds of  ${\boldsymbol{\eta}}_{DHT}$ are computed for $\omega=2$
and $t=1$  thereby obtaining the output displayed in Figure \ref{figrotduff}a). These manifolds have been obtained with the algorithm  reported in Subsection 4.2 which follows the approach by \cite{mani,nlpg}.
 The figure confirms that manifolds are objects rotating with the coordinates.  Figure \ref{figrotduff}b) confirms that the Lagrangian descriptors  discussed in Section 3
provide the same manifolds in the rotated frame.

A second use of the function $M$ is the  computation {\it limit coordinates} that are at the basis of the definition of {\it distinguished trajectory} given in \cite{chaos}. In this paper the authors  have shown that the definition of DT discussed in 4.1 is  robust with respect to rotations in the sense that  in the rotating frame the expression ${\boldsymbol{\eta}}_{DHT}$  is equally characterized as Distinguished.

The linear example in section 4.1  also illustrates that   translations with constant velocity   preserve Distinguished Trajectories (DT) in the sense that in the new frame of reference the transformed DT also  preserves the property of being Distinguished. However not all coordinate transformations preserve  
Distinguished Trajectories.  In particular, coordinate transformations involving trajectories of the velocity field do not necessarily preserve fixed points or their stable and unstable manifolds, and we illustrate this next.

Let us consider the dynamical system:
\begin{eqnarray}
\frac{dx}{dt} &=& x-1 \nonumber \\ 
\frac{dy}{dt} &=& -y \label{expf}
\end{eqnarray}

\noindent
for which $(x=1,y=0)$ is a  hyperbolic fixed point. Let us consider  the trajectory ${\bf x}^p(t)$:
\begin{equation}
{\bf x}^p(t)= \binom{1.5 {\rm e}^{t}+1}{0.5{\rm e}^{-t}}.
\end{equation}

\noindent
  A coordinate transformation based on this trajectory is the following: 
  \begin{equation}
  {\bf x}^N={\bf x}-{\bf x}^p(t), \label{trtr}
 \end{equation}
 
 \noindent
 which  transforms the system (\ref{expf}) into:
  \begin{eqnarray}
\frac{dx^N}{dt} &=& x^N, \nonumber \\ 
\frac{dy^N}{dt} &=& -y^N. \label{expfn}
\end{eqnarray}

\noindent
This is again an autonomous dynamical system with a hyperbolic fixed point at $(x^N=0,y^N=0)$.  The time dependent coordinate transformation (\ref{trtr})  obviously does not transform this fixed point into the old one $(x=1,y=0)$.  Moreover this transformation does not preserve the stable and unstable manifolds themselves. 
The unstable manifold of the fixed point $(x, y) = (1, 0)$  in the original system is given by:
\begin{equation}
{\bf x}_u= \binom{\alpha+1 }{0}
\end{equation}

\noindent
for arbitrary real $\alpha \ne 0$. On the other hand the unstable manifold of the hyperbolic fixed point $(x, y) = (0, 0)$ in the  transformed system is given by:
\begin{equation}
{\bf x}^N_u= \binom{\beta }{0}
\end{equation}

\noindent
for arbitrary real  $\beta \ne 0$.  The transformation   (\ref{trtr}) does not map 
a point in  the unstable subspace ${\bf x}_u$ to a point
 in the unstable subspace ${\bf x}^N_u$ since in general:
\begin{equation}
 \binom{\beta }{0} \neq  \binom{\alpha +1}{0} -\binom{1.5 {\rm e}^{t}+1}{0.5{\rm e}^{-t}}.
\end{equation}

We note that time-dependent transformations based on trajectories may have even more dramatic effects on invariant objects, such as tori. For example, if 
${\bf x}^p(t)$ corresponds to a trajectory  in a torus in the original system it will transform to a fixed point under this transformation.

 Finally, we consider another example of a transformation of coordinates based on trajectories of the original system.
Consider  the one-dimensional autonomous system:
 \begin{equation}
 \frac{dx}{dt}=c \label{trex}
 \end{equation}
 \noindent
where $c$ is a nonzero constant  constant. This system has no fixed points  However if we consider the  transformation (\ref{trtr}) based on any 
 trajectory of  (\ref{trex}) ${\bf x}^p(t)=ct+d$:
   \begin{equation}
  {\bf x}^N={\bf x}-(ct+d), \label{trtrg}
 \end{equation}
 \noindent
  the system becomes: 
 \begin{equation}
 \frac{dx^N}{dt}=0 
 \end{equation}
 \noindent
which is also autonomous and all initial conditions are fixed points. We note that the reason that the transformation  (\ref{trtrg})  does not preserve fixed points is  that it is based on a trajectory of the original system, despite the fact that it is a Galilean transformation.

Finally, we remark that it has been often stated that Lagrangian ``structures'' and analytical methods  should be frame-independent (see for instance  \cite{mh12}).  However, from these examples we see
that fixed points and invariant manifolds of hyperbolic fixed points  may not be preserved by  transformations based on particle trajectories. 
This indicates that more reflection is required on what is meant in this context by frame-independence and what truly must be demanded of geometrical structures and analytical tools for  useful Lagrangian descriptions.

 \section{Transport routes across the ocean surface}
 
\begin{figure*}
a)\includegraphics[width=6.cm]{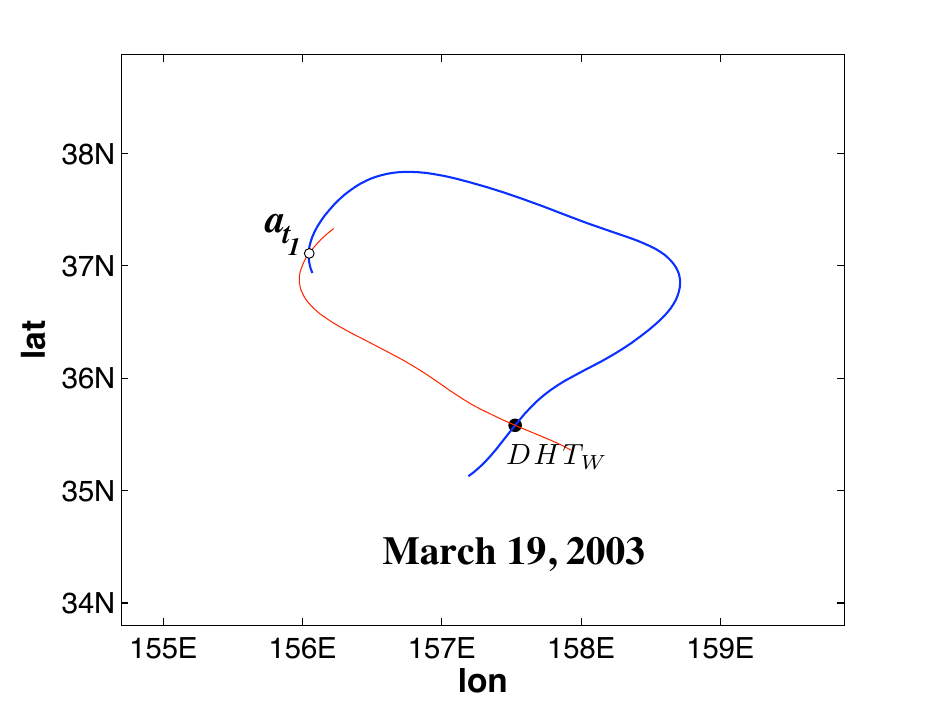}b)\includegraphics[width=6.cm]{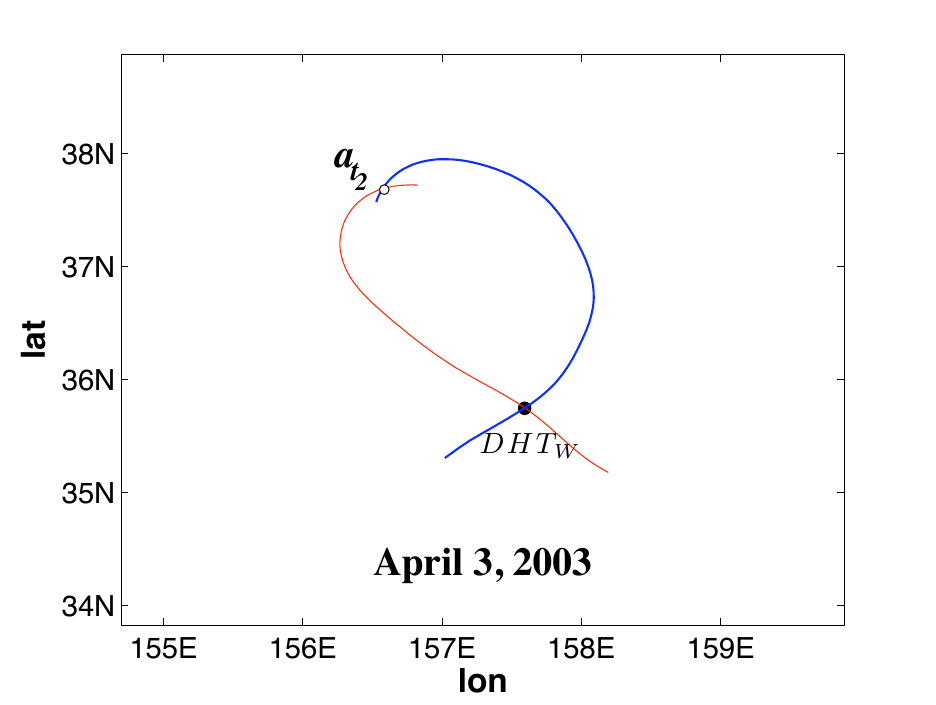}
\caption{\label{barriere}Lagrangian barriers for the western eddy at dates March 19 and  April 3, 2003. These have been made from  finite length pieces of the stable and unstable manifolds of DHT$_W$. The boundary intersection points
are denoted respectively by $a_{t_1}$ and $a_{t_2}$. }
\end{figure*}

\begin{figure*}
a)\includegraphics[width=7cm]{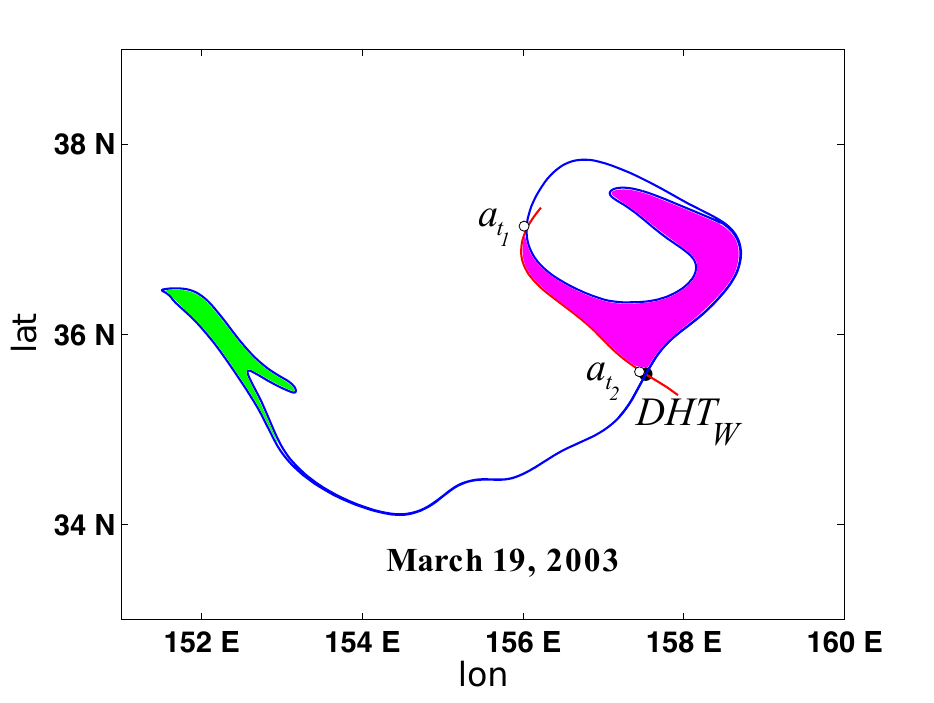}b)\includegraphics[width=7cm]{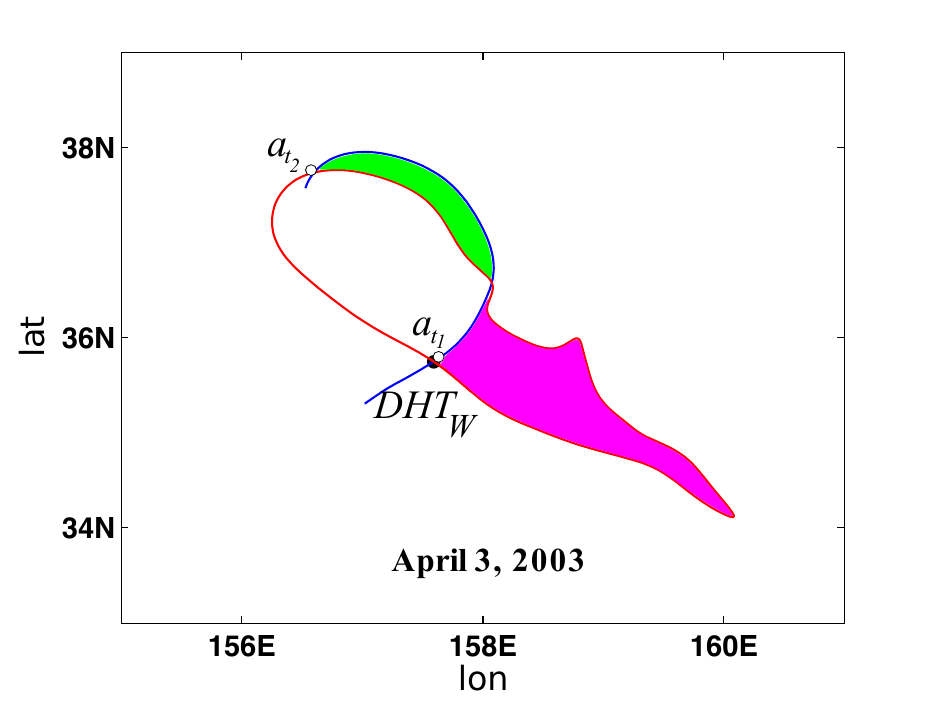}
\caption{\label{turnstilee} Turnstile lobes across the western eddy at dates March 19 and April 3, 2003. The intersection trajectories $a_{t_1}$ and $a_{t_2}$ are displayed at both dates
showing their clockwise circulation around the eddy. The magenta area evolves from the inside to the outside while the green area does  from the outside to the inside.  }
\end{figure*}

\begin{figure*}
a)\includegraphics[width=7cm]{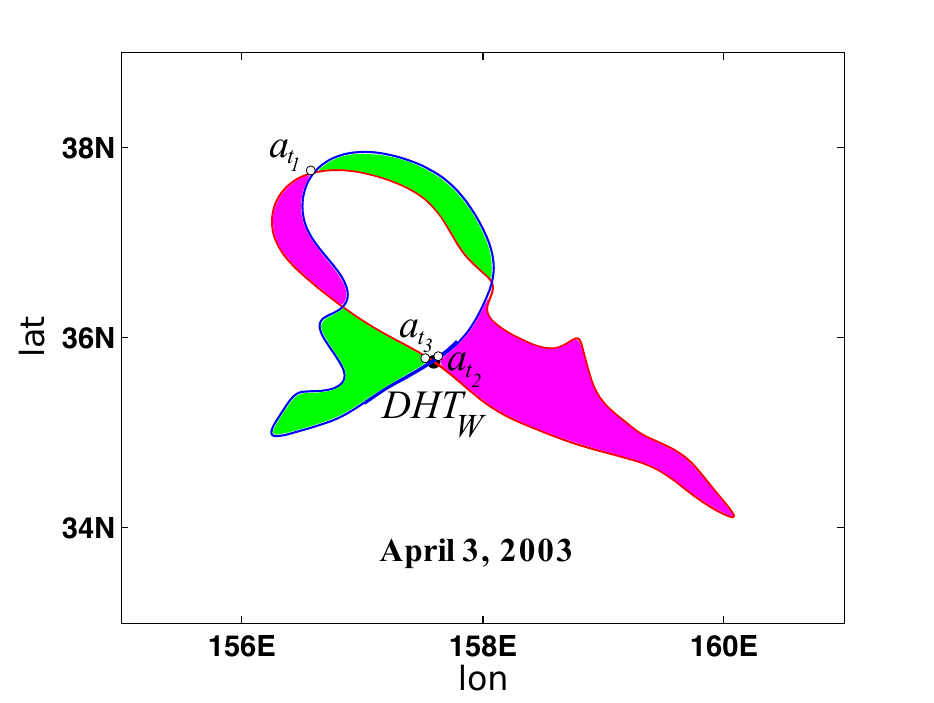}b)\includegraphics[width=7cm]{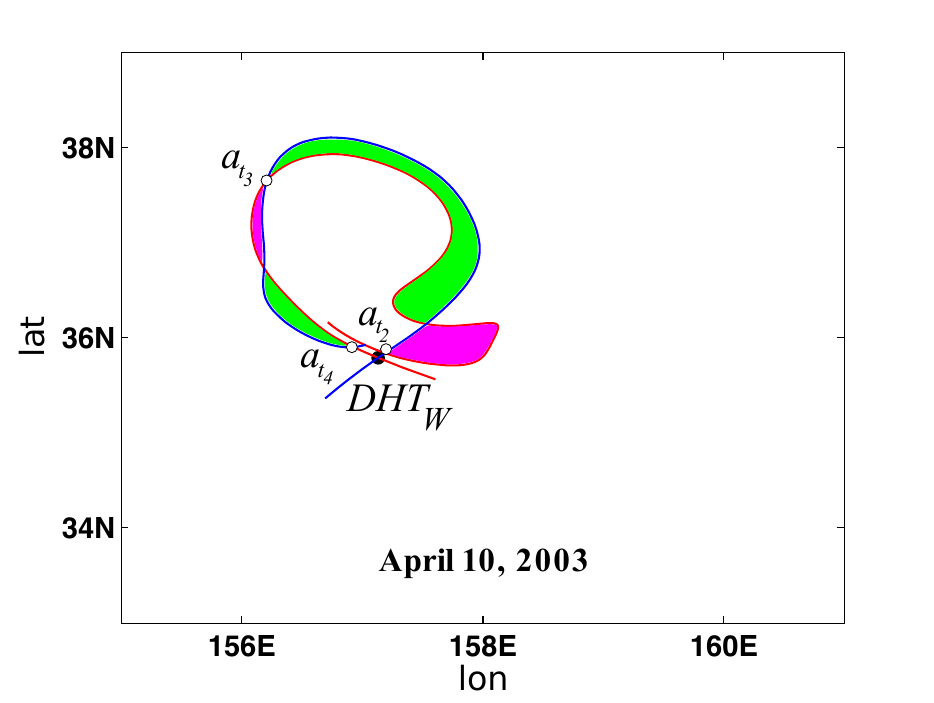}\\c)\includegraphics[width=7cm]{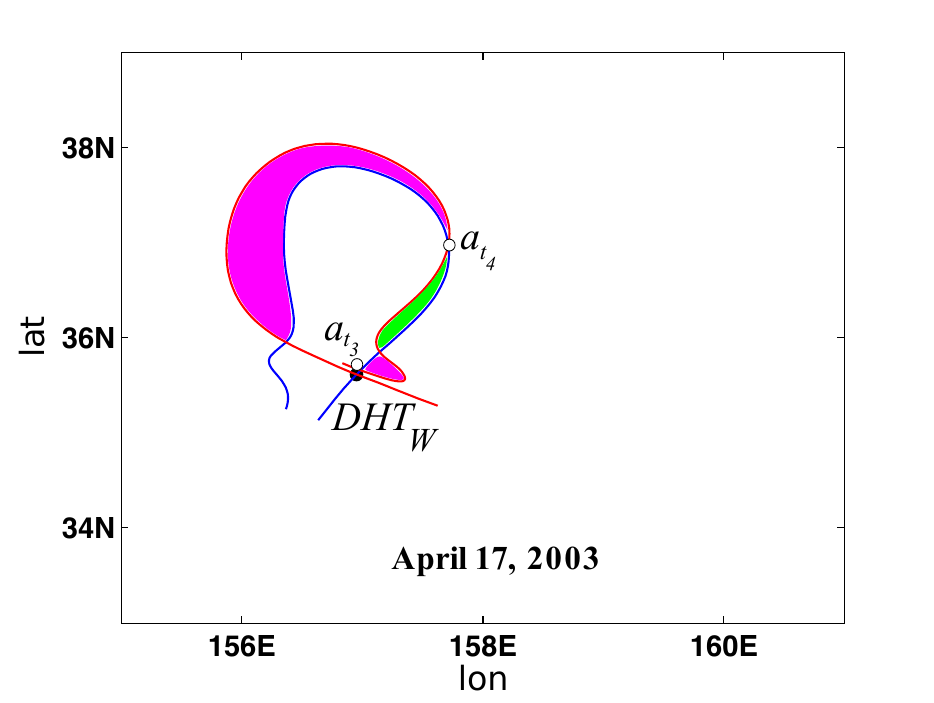}d)\includegraphics[width=7cm]{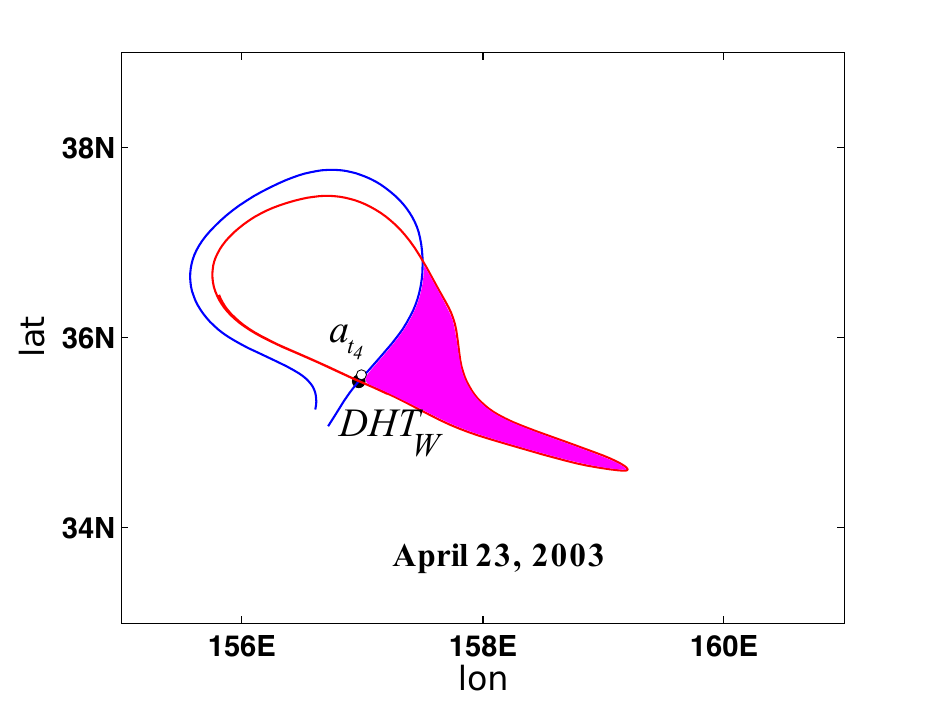}
\caption{\label{sequencee} Sequence of lobes mixing waters from inside the eddy to outside and viceversa in selected days of year 2003. a) April 3; b) April 10; c) April 17; d) April 23.  }
\end{figure*}

 In this section we show how to obtain transport information from the output of the tools  described
in previous sections. We  start by describing transport  across  eddies   displayed in 
Figure  \ref{fig:M1}.
Particles in their interior, despite belonging to flows in a quite chaotic regime,  as is the case of the ocean surface,  typically do not    experience the butterfly effect which is
 characterised by 
 a high sensitivity to initial conditions. On the contrary, particles contained therein remain gathered together for long periods of time, during which they form spatially {\it coherent structures}.
 Mathematically, eddies are related to non-hyperbolic flow regions, where particles evolve mostly ``circling".
 The exponentially increasing separation between particles is   characteristic of hyperbolic regions, which are also responsible for 
 unpredictability. Essentially transport across the ocean surface  
 is governed by the interplay between these dispersive and non dispersive objects. 
The Lagrangian description of eddies 
  identifies the existence of  an outer collar, where the interchange with the media is understood in terms of lobe dynamics 
  (see \cite{bernard,bra}) and an inner core, which is robust and rather impermeable to stirring, as already described
   in section 3. In this section we focus on
   describing  transport across the outer part of the eddy which is located at the west end in Figure \ref{fig:M1}. The stable and unstable 
   manifolds  of $DHT_W$, which are involved in the transport across this vortex, are shown  in  Figure \ref{mani}. These manifolds confirm the exchange
   of water by the presence of the turnstile mechanism
 for a period of one month from March 19 to April 23. The turnstile  mechanism has been extensively used and explained in the literature \citep{maw,rlw},  and has been found 
to play a role in transport in several oceanographic contexts \citep{nlpg2,jpo,cw}. 
This mechanism is described  from pieces of stable and unstable 
manifolds of the identified  DHT. A first point 
to address is the selection of those pieces of invariant manifolds from messy curves such  as those in Figure \ref{mani}. 
For this purpose we consider  that a manifold has two branches separated by the DHT which is taken as a reference point on 
the manifold, and selections of portions of manifolds are made from this reference point. 
Given that trajectories may retain the distinguished property only in finite time intervals,  the 
identification of the two branches on the manifold is possible only on time intervals when the trajectory remains distinguished. Beyond that time 
the manifold computation may continue, but the  reference point on it is lost. The turnstile mechanism
identifies masses of water crossing a time-dependent Lagrangian barrier   separating the inside from the outside. 
 The Lagrangian barrier around the  vortex in Fig \ref{mani} at a time $t_k$  is defined by selecting a branch of the unstable manifold which starts at  DHT$_W$ and surrounds the eddy towards the left side
  and  a branch of the stable  manifold which starts at DHT$_W$ and surrounds the eddy towards the right side. We choose the segments considering that they must intersect at precisely one point $a_{t_k}$  and
  that they must form a relatively smooth boundary (i.e, free of the violent oscillations displayed by each of the manifolds when approaching  DHT$_W$ from the opposite side).
 Figure \ref{barriere} shows the selections  outlining the barriers   
for the dates $t_1=$March 19 and  $t_2=$April 3, 2003. The blue line stands for the stable manifold while the red line corresponds  to the unstable manifold. 
The boundary intersection points are marked
as $a_{t_1}$ and $a_{t_2}$.  Intersection points are  invariant, which means that  if the stable and unstable manifolds
intersect  in a point at a given time, then they intersect for all time, and the intersection point is hence a trajectory. 
For a better  understanding of  the time evolution of lobes, the positions of 
trajectories $a_{t_1}$ and $a_{t_2}$ are depicted at different times.  
Figure \ref{turnstilee} shows longer pieces of the unstable and stable manifolds at the same days as those selected in Figure \ref{barriere}. 
Manifolds intersect  forming regions called lobes. Only the fluid that is inside the lobes  can participate in
the turnstile mechanism. Two snapshots showing the evolution of  lobes from 
March 19 to April 3 are displayed. There one may observe how the lobe which is inside the eddy on March 19
goes outside on April 3. Similarly
the lobe which is outside on March 19 is inside on April 3. Trajectories $a_{t_1}$ and $a_{t_2}$ are depicted, showing that 
they evolve, circulating   clockwise around the DHT$_W$.  The green colour applies to the lobe  that evolves towards the interior of the eddy while 
the magenta area evolves from the inside towards the outside.
Between March 19 and April 23,   several lobes are formed, mixing waters at both sides of the eddy.  Figure \ref{sequencee} contains a time sequence showing the evolution 
of several lobes created by the intersection of the stable and unstable manifolds. The selected days are:  $t_2=$April 3,  $t_3=$April 10, $t_4=$ April 17 and $t_5=$ April 23. A sequence of trajectories $a_{t_1},a_{t_2},a_{t_3}, ...$ 
obtained from the intersection points is depicted. These trajectories evolve clockwise, surrounding the vortex,   and serve as references for tracking lobe evolution. 
Beyond April 23  we cannot locate further intersections between the stable and unstable manifolds of DHT$_W$. Hence, no more lobes are found, and our description of  the turnstile 
mechanism ceases. 

\begin{figure*}[t]
\vspace*{2mm}
\begin{center}
\includegraphics[width=8cm]{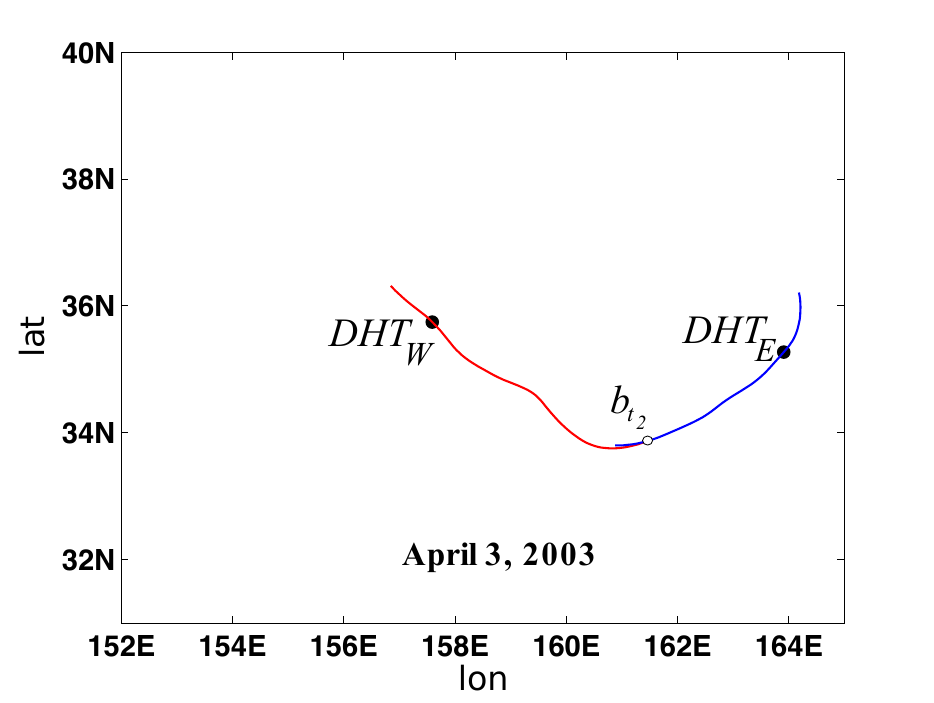}\includegraphics[width=8cm]{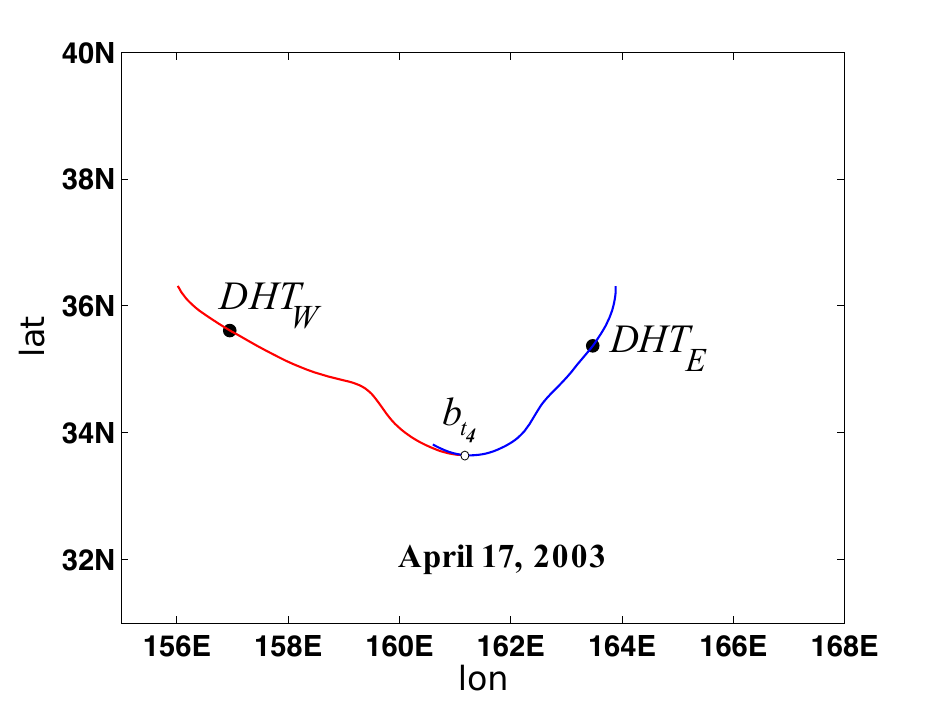}
\end{center}
\caption{Boundaries at days April 3 2003 and April 17 2003 constructed from a (finite length) segment of the
unstable manifold of DHT$_W$ and a (finite length) segment of the stable manifold of DHT$_E$. The boundary intersection points are denoted respectively by $b_{t_2}$ and $b_{t_4}$. (Figure taken from \cite{nlpg2}). }
\label{barrier}
\end{figure*}

\begin{figure*}
a)\includegraphics[width=7cm]{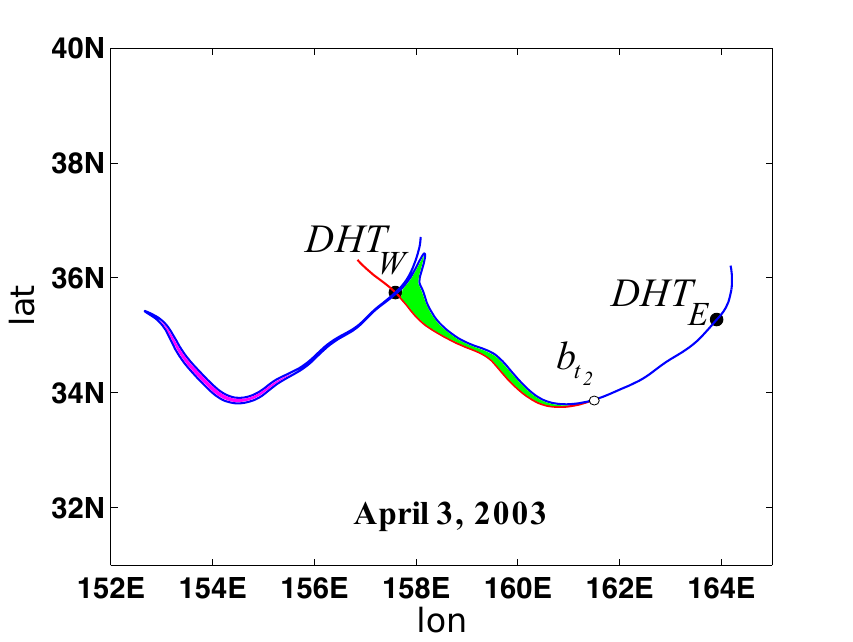}b)\includegraphics[width=7cm]{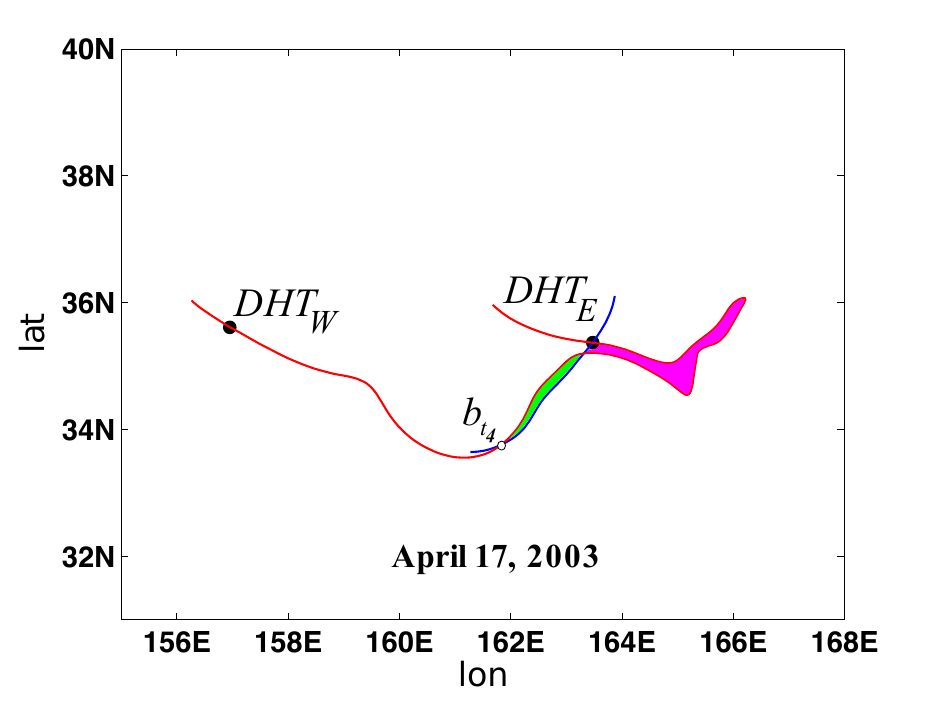}
\caption{\label{sumnpg}Turnstile lobes across the main stream at days April 3  and 17 2003.  The magenta area evolves from south to north  while the green area does  from north to south. (Figure taken from \cite{nlpg2}).}
\end{figure*}

The turnstile mechanism across the eddy coexists in time with other transport  routes observed, for instance, across structures such as the   reddish main  current    in Figure  \ref{fig:M1}.
\cite{nlpg2} have addressed transport across this jet  in terms of DHT and invariant manifolds.
 There it  has been found that the turnstile mechanism is active in transporting masses of water across such a current, and it
 has been proven that the exchange survives between 3 April 2003 and 26 May 2003.
To provide a complete overview of the whole transport picture, we next summarise  the results reported by \citep{nlpg2}.  
The turnstile mechanism is described  from pieces of stable and unstable 
manifolds of the identified  DHTs,  at the east and west limits of the main stream. The  mechanism
identifies masses of water crossing the time dependent Lagrangian barriers  depicted in Fig. \ref{barrier}, which separates north from south. 
The figure  shows
 a piece of the unstable manifold of DHT$_W$ and a piece of the stable  manifold of DHT$_E$ that define those barriers on days {April 3 and April 17}. For consistency with the notation used to describe transport
across the eddy we name these dates as $t_2=$April 3 and $t_4=$ April 17.
Only portions of one branch are displayed for each DHT.
 They intersect at points  marked with letters $b_{t_2}$ and $b_{t_4}$.  They are trajectories which maintain their labels in all pictures in order for  the lobe evolution to be easily tracked.
Longer pieces of the same manifolds are represented in
 figure \ref{sumnpg}. Figure \ref{sumnpg}b) shows  the asymptotic evolution on April 17
of the lobes represented in Figure \ref{sumnpg}a) on April 3. The green lobe area  contains particles in the north 
on April 3 that eventually came to the south on April 17.  Magenta particles 
that are analogously first  in the south  eventually  come to the north on April 17. 
Lobe dynamics across the main stream may be identified until  26 May 2003. On this date, DHT$_W$ has lost its distinguished property
and the reference point on the unstable manifold has disappeared.  \cite{nlpg2} have reported that 
it is possible  to identify a new reference point on the manifold, which is given
by  DHT$_W^+$.  The manifold is not asymptotic to DHT$_W^+$.  However, DHT$_W^+$  marks a Distinguished Trajectory on the manifold with certain accuracy $\epsilon$.

\begin{figure*}
\includegraphics[width=12.5cm]{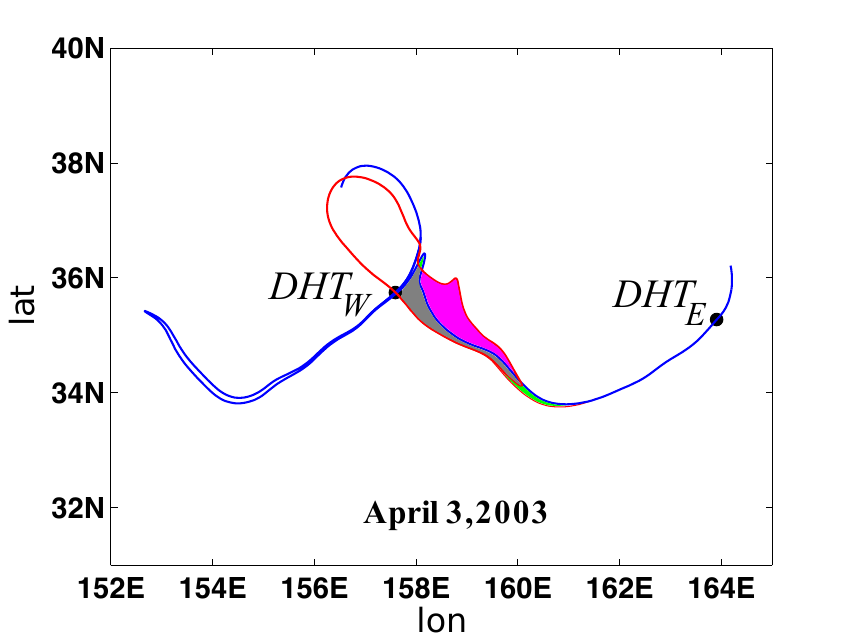}
\caption{\label{inters} Intersection of lobes governing transport across the western eddy and those governing transport across the main stream of the Kuroshio current.}
\end{figure*}

The active transport mechanisms just described are simultaneous in time and the full description of transport routes
should address how their action over  ocean particles is combined.
A complete representation   of coincident events in Figure  \ref{inters} reveals  intersections 
between the lobe that  is outside the eddy (magenta colour in Fig. 20a) on April 3, and the lobe which  
at the same time is located to the north of the barrier (green colour in Fig. 22a). 
The intersection  area in grey colour, as shown in Figure  \ref{inters} for April 3,  provides dual information on the particles  contained therein.
It shows that those particles  were inside the eddy on March 19 (as indicated in Figure \ref{turnstilee} ) and were to be at the south of the Lagrangian barrier 
across the stream on April 17 (see Figure \ref{sumnpg}).
Further similar intersections take place between the magenta lobes  in  the sequence displayed in Figure  \ref{sequencee}, and 
the sequence of lobes across the jet  that transports water from north to south (see  \cite{nlpg2} for a full  description). Once particles reach the southern region,
 further interactions will take place with 
any dynamic structure covering  the ocean surface in that area.

\begin{figure*}
\includegraphics[width=14.cm]{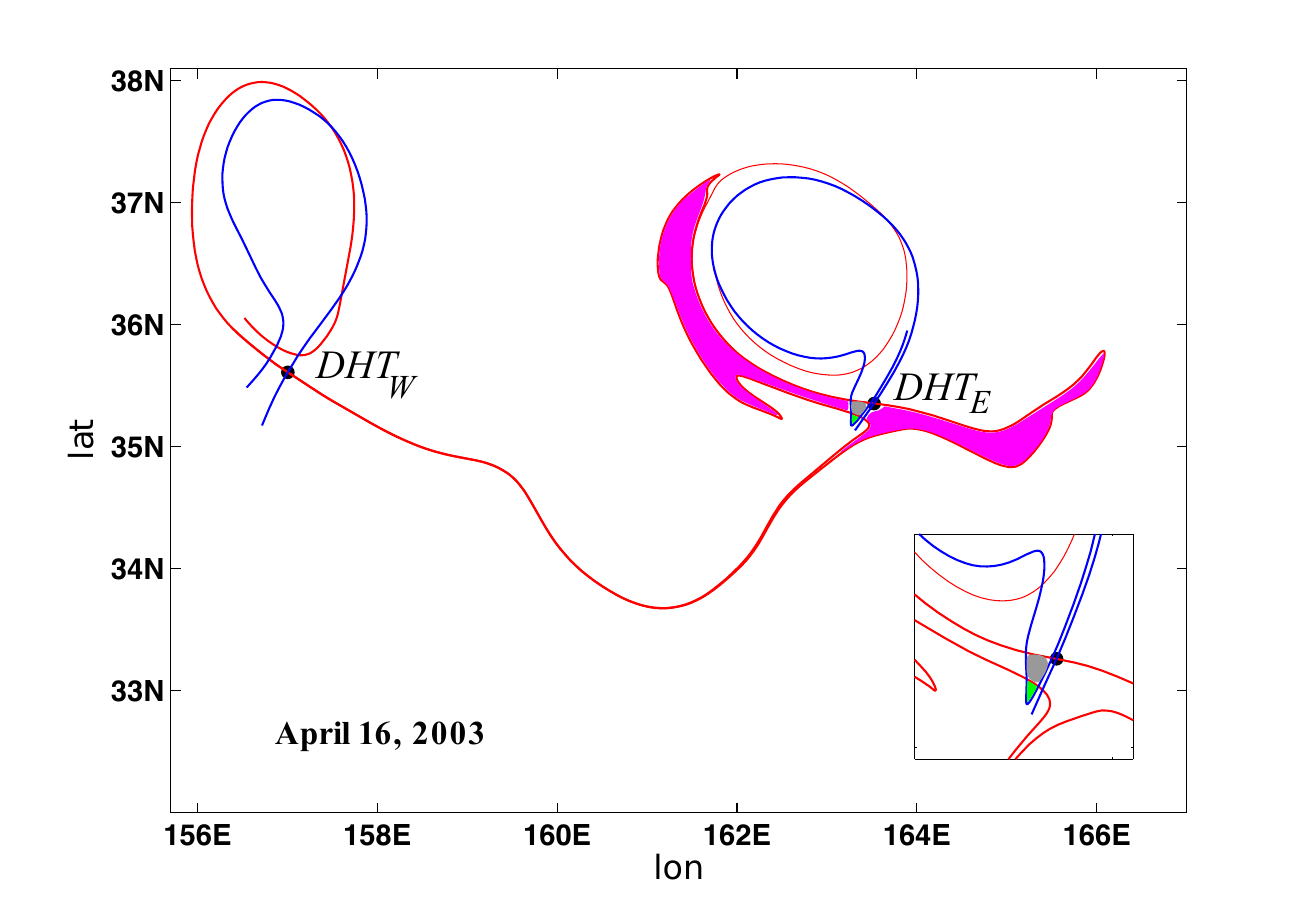}\\\includegraphics[width=14.cm]{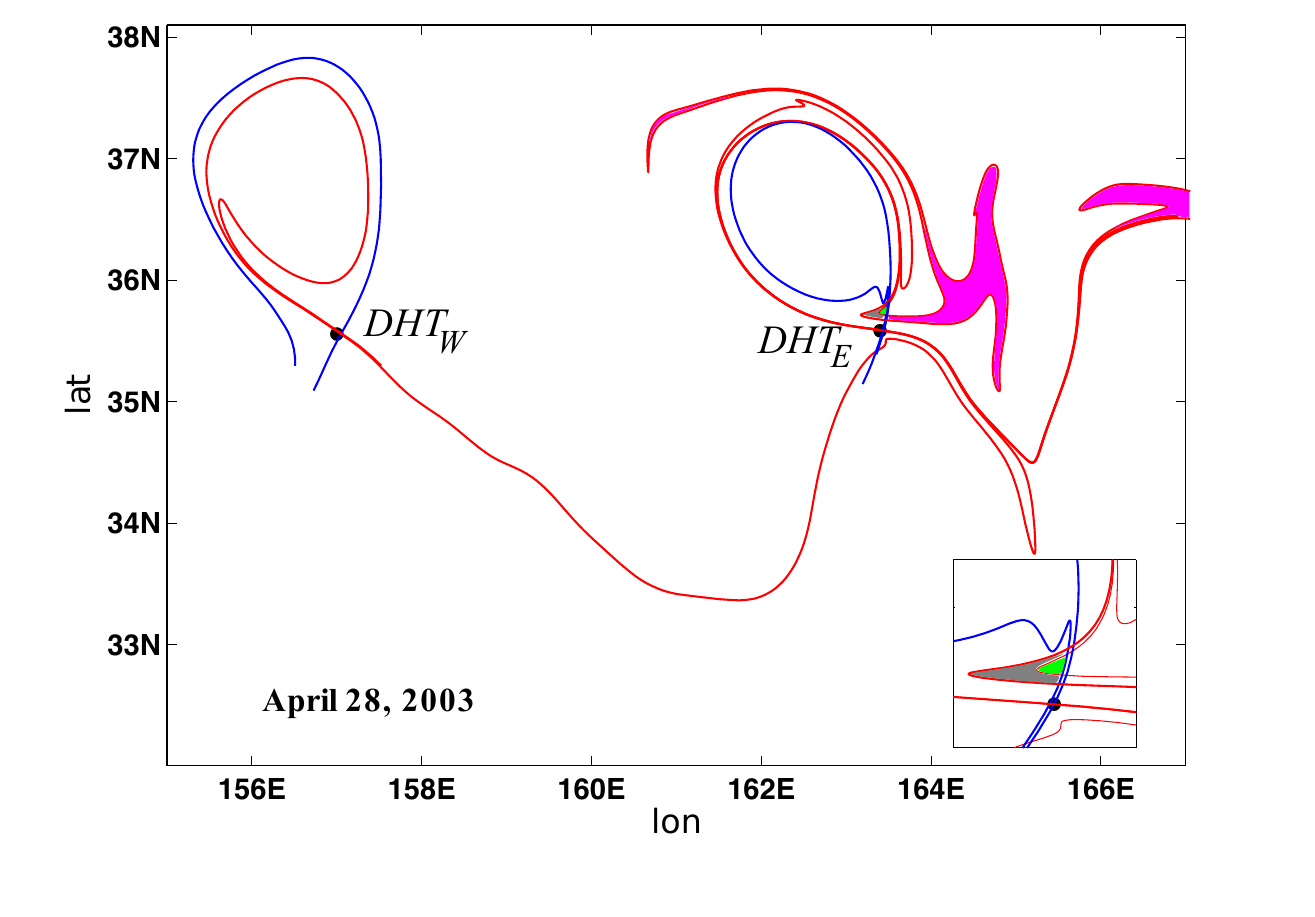}
\caption{\label{intersede}Intersection of lobes governing transport across the western and eastern eddies. a) On April 16 the  grey 
area shows a portion of fluid ejected from the western eddy that will be entrained by the eastern eddy; b) on April 28 the grey area has come into the eastern eddy. Figures show insets with an amplification of the entrainment process.}
\end{figure*}

Additional complex routes may be traced for particles ejected from the western eddy. In fact, we are able to show that
 there is a non-zero flux from this eddy  towards the eddy at the eastern limit. 
On April 16, Figure \ref{intersede}a)  shows pieces of stable and unstable manifolds of the eddies at the west and east. The magenta coloured lobe represents the water ejected from the western eddy.
There exists  a non-zero intersection area  between this lobe and
the lobe regulating the water coming into the eastern eddy. The intersection area is depicted in dark grey. A remaining piece of the lobe penetrating on the eastern eddy is left in green.  Figure \ref{intersede} b) confirms the entrainment of this area on the eastern vortex on April  28.

A complete transport description would require connection of  the information provided by all the dynamic structures tilling the ocean surface
which are displayed by the function $M$ in Figure \ref{fig:M1}. 
However, in practice, providing thorough insights in terms of manifolds  as  discussed in this section is not always possible, because on the one hand it requires 
 that the features of the observed dynamic patterns resemble those described in the mathematical literature, and
on the other that they  must have certain persistence in time. Rapidly transient regimes,  such  as those occurring in large areas of the analysed  flow,
are difficult to understand  because they are related to changes in the topology of the flow that are not well interpreted
from a dynamic point of view (see \cite{jpo}).  Related to these changes are  mathematical issues such as non-uniform hyperbolicity, addressed for instance in \citep{barreiro}, not yet 
completely understood for general non autonomous  systems such as those represented by Eq. (\ref{adv}).

\section{Conclusions}

This article reports  the combined used  of several Lagrangian tools, some of them recently developed,  and shows  their success in obtaining extensive details about the description
of purely advective transport events in arbitrary time dependent flows. We demonstrate the capabilities of these tools by analysing 2D data sets obtained from altimetric satellites
over the Kuroshio Current.

We have first considered the evaluation of  global Lagrangian descriptors over a general vector field. In particular we have chosen two types of descriptors, referred to as function $M$.
 Contour plots of these functions  provide a time dependent phase portrait which is visualised by sharp changes in the colour code of $M$.
These abrupt variations separate regions of trajectories with qualitatively different behaviours, and since this is exactly what invariant  manifolds separate, boundaries of homogenous coloured areas  
position invariant manifolds.  The dynamic  picture provided by $M$  reveals at a glance the organising centres of the flow, hyperbolic and non-hyperbolic flow regions, invariant manifolds and jets.
In other words it identifies the essential dynamical elements that must be considered by any  kinematic model describing the exchange of trajectories on a given data set. 
 
 Although  the dynamical structures are clearly  visualised  from $M$, a detailed description of transport requires the full identification of the organising trajectories,  the Distinguished 
 Hyperbolic Trajectories, and of their finite time stable and unstable manifolds. Our discussions are focused on 2D flows, although extension to higher dimensions are possible.  Distinguished  hyperbolic  trajectories are computed by first examining $M$ as defined from   Eq. (\ref{def:Mgen}), and  identifying candidate 
 areas which  act as the organising centres of the flow.  The search is completed 
 by computing paths of limit coordinates 
 on each recognised area for a   full identification of the DHT positions.
 At a  third  stage, finite time stable and unstable manifolds of these DHTs are directly computed as advected curves. The algorithm starts 
with a small segment aligned either along the stable or the unstable subspace of the DHT, making  this segment evolve either backwards or forwards in time respectively. 
Manifolds computed in this way become long intricate curves;  transport details are obtained from them by selecting portions 
along the branches  at  both sides of the DHT. These selections allow  transport routes across the ocean surface to be identified; for instance, masses of water  penetrating or leaving an eddy,
then of those masses protruding the eddy, parcels are identified  crossing the main current or coming into a second eddy.  A complete transport description 
 connecting the information provided by all the dynamic structures tilling the ocean surface is foreseen.
Despite the advances made,  however  a full transport description  still remains a challenge  because conceptual difficulties  exist that are yet to be solved, 
especially when dealing with highly transient regimes in which the topology of the flow changes in time.

As a summary, we can say  that our Lagrangian techniques  have proven fluid exchange across the main current and between eddies  in the Kuroshio region  in a range of dates during the year 2003.
This methodology constitutes an efficient  tool of analysis for the uncountable data sets which  nowadays are obtained  from altimeter satellite or by other means.
The performance of the machinery on the analyzed data  opens a gateway to its applications in any kind of realistic flow for operational oceanography   purposes and could be though 
as an alternative for the study of transport in oceanic flows to campaign measures based on quasi Lagrangian drifter releases.

\section*{Acknowledgements}
We are indebted with S. Wiggins for his very insightful suggestions.  We also acknowledge J. Porter for his comments.
The computational part of this work was done using the CESGA computer FINIS TERRAE and computers at Centro de Computacion Cientifica (UAM).
The authors have been supported by CSIC Grants OCEANTECH No. PIF06-059 and ILINK-0145, Consolider
I-MATH C3-0104, MINECO Grants Nos.   MTM2011-26696 and ICMAT Severo Ochoa project SEV-2011-0087 and the Comunidad de Madrid
Project No. SIMUMAT S-0505-ESP-0158.

%

\bibliographystyle{jfm}

\bibliography{jfm2esam}
\section*{Appendix A}

We   discuss  here details about the numerical evaluation of
$M$ as defined in Eq. (\ref{def:Mgen}).  Trajectories $(x_1(t), x_2(t))$ of the system (\ref{sd1def})-(\ref{sd2def}) are obtained numerically,
and thus represented by  a finite number of points, $L$. A discrete version of Eq. (\ref{def:Mgen}) is:
\begin{equation}
M= \sum_{j=1}^{L-1} \left(  \int^{p_f}_{p_i} \! \!\!\sqrt{\left(\frac{d x_{1,j}(p)}{dp}\right)^2+\left(\frac{d x_{2,j}(p)}{dp}\right)^2 }dp \right), 
\end{equation}
where the functions $x_{1,j}(p)$ and $x_{2,j}(p)$ represent a curve interpolation parametrised by $p$, and the integral
\begin{equation}
\int^{p_f}_{p_i} \! \!\!\sqrt{\left(\frac{d x_{1,j}(p)}{dp}\right)^2+\left(\frac{d x_{2,j}(p)}{dp}\right)^2 }dp \label{integral}
\end{equation}
is computed  numerically. In accordance with the methodology described in \citep{chaos}, we  use
 the interpolation method proposed by  \cite{dr} in the context of contour dynamics. The interpolation equation,  later used in this article, is given by expression (\ref{eq:drint}). To compute the integral (\ref{integral})
we have used the Romberges method (see \cite{nr}) of  order $2K$ where $K=5$. In the results reported in this article we have used this technique to evaluate  Eq. (\ref{def:Mgen}).  Another possibility for evaluating Eq. (\ref{integral}), which is less accurate but simpler and faster, is to approach it by the length of the linear 
segments linking successive points of the trajectory. 
In order to evaluate  Eq. (\ref{def:M})
where  $\mathcal{F}({\bf x}(t))$ depends not only on velocity but also  on other vectors
  such as  acceleration, the time derivative of acceleration or their combinations, we propose a more versatile method which is easily adapted for   any choice of $\mathcal{F}$. 
For instance, in the case where    $\mathcal{F}({\bf x}(t))=|{\bf v}({\bf x}(t),t)|$, the integral in Eq. (\ref{def:M}) evaluates the area $A$ below the graph $|{\bf v}({\bf x}(t),t)|$ in the referred time interval. 
In order to evaluate $A$ we consider the integral as the following one-dimensional dynamical system:
\begin{equation}
\frac{d Y}{dt}=|{\bf v}({\bf x}(t),t)|. \label{1dds}
\end{equation}
For the initial condition $Y(t^*)=0$, the  area $A$ is provided  by the value of $Y$ at $t^*+\tau$ minus  the value of $Y$ at $t^*-\tau$ ,  i.e.,   $Y(t^*+\tau)-Y(t^*-\tau)=A$.
The integration of the system (\ref{1dds}) is performed with a 5th order variable time step Runge-Kutta method, in particular with the subroutine {\tt rkqs} described in \cite{nr}.
The peculiarity of this differential equation  is that it depends on $t$ both explicitly and implicitly (through the trajectory), and expressions such as the right hand side
of  the system (\ref{sd1def})-(\ref{sd2def})  only provide the explicit dependence  ${\bf v}({\bf x},t)$.
A Runge Kutta step from $t_0$ to $t_1$ applied to Eq. (\ref{1dds}) requires the evaluation of $|{\bf v}|$ along trajectories at intermediate steps $t_0+\Delta t$. To this end
the argument ${\bf x}$ that must be passed to  $|{\bf v}|$ at time $t_0+\Delta t$
must be obtained by evolving the trajectory from ($t_0, {\bf x}(t_0)$) to ($t_0+\Delta t, {\bf x}(t_0+\Delta t)$) according to   the system (\ref{sd1def})-(\ref{sd2def}).
This method
  is quite adaptable, since  from one descriptor to another it is only the right hand side in Eq. (\ref{1dds}) which needs to be modified. This
  is the  technique used for the case in which  $\mathcal{F}({\bf x}(t))=|{\bf a}({\bf x}(t),t)|$, for which we report results.
  
\section*{Appendix B}
We provide full details of the equations and algorithms used to compute the unstable manifolds. 
At each time $t_k$ in a discrete set of time increments $[t_k, k=0...N]$,  the unstable manifold  is represented by a discrete set of points ${\bf x}_j$. In particular at time $t_0$ it 
is a small segment aligned along the unstable subspace of the hyperbolic trajectory, represented by five points. They are evolved along trajectories until time $t_1$, and each point is considered
to leave unacceptable gaps  with its neighbors if the measure $\sigma_j>1$.  Here $\sigma_j=d_j \rho_j$ where   $d_j= {\bf x}_{j+1}-{\bf x}_{j}$ and $\rho_j$ is a density  defined as follows:
\begin{equation}
\rho_j \equiv \frac{(\bar{\kappa}_j L)^\frac{1}{2}}{\mu L} + 
\bar{\kappa}_j,\label{eq:rho}
\end{equation}
or $2/\zeta$, whichever is smaller. Here $\zeta$ serves as a small-scale cut-off 
distance for resolving manifold details which we have fixed to $10^{-6}$ and $L$ is a typical length scale fixed to 3. 
  The parameter $\mu$ 
controls the overall point density along the manifold and needs
tuning for individual problems.  Small values
of $\mu$ correspond to a high point density. In our computations it is fixed to 0.005.
The quantity $\bar{\kappa}_j$ in (\ref{eq:rho}) is defined in terms of
$\check{\kappa}$,
\begin{align}
{\bar{\kappa}_j} &\equiv (\check{\kappa}_j+\check{\kappa}_{j+1})/2, \\
\intertext{which in turn is defined by}
\check{\kappa}_j &= \frac{w_{j-1} \tilde{\kappa}_{j-1} 
  + w_{j} \tilde{\kappa}_{j}}{w_{j-1}+w_{j}},
\intertext{which uses the weighting $w_j=d_j/(d_j^2+4\zeta^2)$ and the 
further curvature $\tilde{\kappa}_{j}$, which itself is defined by}
\tilde{\kappa}_{j} &= \sqrt{\kappa_j^2 +1/L^2}, 
\end{align}
where $\kappa_j$, finally, is the local curvature:
\begin{equation}
\kappa_j = 2\frac{a_{j-1}b_j-b_{j-1}a_j}{|d_{j-1}^2  {\mathbf t}_j+d_{j}^2 {\mathbf t}_{j-1} |} .\label{eq:kappa}
\end{equation}
Here 
\begin{alignat}{2}
{\mathbf t}_j &= (a_j,b_j)={{\mathbf  x}}_{j+1}-{{\mathbf  x}}_j, &&  \qquad{t}_j\in {\mathbb R}^{2}\label{eq:t}
\end{alignat}

When a gap between nodes at time $t_1$ is too large, it is filled by inserting a point  between the same nodes at $t_0$.  The point is computed by interpolating with $p=0.5$ along the 
curve  that links the points ${\bf x}_{j+1}, {\bf x}_{j}$:  
\begin{equation}
\label{eq:drint}
{\mathbf x}(p) = {\mathbf x}_j + p~{\mathbf t}_j+ {\eta}_j(p)~{\mathbf n}_j,
\end{equation}
where $ {\mathbf t}_j $ is given by Eq. (\ref{eq:t}) and:
\begin{alignat}{2}
{\mathbf n}_j &=  (-b_j, a_j),                      &&  \qquad {n}_j\in {\mathbb R}^{2}\\ 
{\eta}_j (p) &= \mu_j p+\beta_j p^2 +\gamma_j p^3, && \qquad {\eta}_j (p) \in{\mathbb R}.
\end{alignat}
The cubic interpolation coefficients 
$\mu_j$, $\beta_j$ and $\gamma_j$ are:
\begin{align}
\mu_j    &=-\frac{1}{3} d_j \kappa_j-\frac{1}{6} d_j \kappa_{j+1},\\
\beta_j  &= \frac{1}{2} d_j \kappa_j,\\
\gamma_j &= \frac{1}{6} d_j (\kappa_{j+1}-\kappa_{j}),
\end{align}

Once the manifold satisfies 
gap size acceptability condition  at every node, {\it i.e.} $d_j \rho_j=\sigma_j<1$,   the point redistribution algorithm is applied. This is useful to eliminate points in
regions of the manifold where they may have accumulated \cite{nlpg}. This algorithm is described
in \citep{dr} and it works as we
describe next.  Let $n$ be the number of nodes at $t_{1}$:
\begin{equation}
q = \sum_{j=1}^{n}\sigma_j
\end{equation}
and define $\tilde{n}=[q]+2$ (i.e., two more than the nearest integer
to $q$).  During redistribution the end points of the manifold are held fixed. The $n-2$ ``old'' nodes between the end points will be
replaced by $\tilde{n}-1$ entirely new nodes in such a way that the
spacing of new nodes is approximately consistent with the desired
average density, controlled by the parameter $\mu$.
Let $\sigma_j'=\sigma_j \tilde{n}/q$ 
so that 
$\sum_{j=1}^{n} \sigma_j'=\tilde{n}$. Then, the
positions of the new nodes $i=2,...,\tilde{n}$ are found succesively
by seeking for each successive $j$ a $p$ such that,
\begin{equation}
 \sum_{l=1}^{j-1} \sigma_l' + \sigma_{j}'p=i-1,
\end{equation}
and placing each new node $i$
between the old nodes $j$ and $j+1$ at the position
${\mathbf x} (p)$ given in (\ref{eq:drint}).

\end{document}